\documentclass[twocolumn]{aastex631}

\newcommand\galform{\texttt{GALFORM} }

\usepackage{graphicx}
\usepackage{enumitem}
\usepackage{appendix}

\begin{document}

\title{Explaining JWST counts with galaxy formation models.}

\author[0000-0001-8220-2324]{Giorgio Manzoni}
\affiliation{Jockey Club Institute for Advanced Study, The Hong Kong University of Science and Technology, Hong Kong S.A.R., China}
\author[0000-0002-8785-8979]{Tom Broadhurst}
\affiliation{Department of Theoretical Physics, University of the Basque Country UPV-EHU, E-48040 Bilbao, Spain}
\affiliation{Donostia International Physics Center (DIPC), 20018 Donostia, The Basque Country, Spain}
\affiliation{IKERBASQUE, Basque Foundation for Science, Alameda Urquijo, 36-5 E-48008 Bilbao, Spain}
\author[0000-0003-4220-2404]{Jeremy Lim}
\affiliation{Department of Physics, The University of Hong Kong, Pokfulam Road, Hong Kong}
\author[0000-0002-5248-5076]{Tao Liu}
\affiliation{Department of Physics, The Hong Kong University of Science and Technology, \\ Clear Water Bay, Kowloon, Hong Kong S.A.R., China}
\affiliation{Jockey Club Institute for Advanced Study, The Hong Kong University of Science and Technology, Hong Kong S.A.R., China}
\author[0000-0001-7575-0816]{George Smoot}
\affiliation{Jockey Club Institute for Advanced Study, The Hong Kong University of Science and Technology, Hong Kong S.A.R., China}
\affiliation{Department of Physics, The Hong Kong University of Science and Technology, \\ Clear Water Bay, Kowloon, Hong Kong S.A.R., China}
\affiliation{DIPC, Basque Country UPV/EHU, E-48080 San Sebastian, Spain}
\affiliation{Energetic Cosmos Laboratory, Nazarbayev University, Nursultan, Kazakhstan}
\affiliation{Physics Department, University of California at Berkeley, Emeritus, Berkeley, California 94720, USA}
\affiliation{Paris Centre for Cosmological Physics, APC, AstroParticule et Cosmologie, Universit\'e de Paris,
CNRS/IN2P3, CEA/lrfu, Universit\'e Sorbonne Paris Cit\'e, 10, rue Alice Domon et Leonie Duquet,
75205 Paris CEDEX 13, France Emeritus}
\author[0000-0002-9935-9755]{Carlton M. Baugh}
\affiliation{Institute for Computational Cosmology (ICC), Department of Physics, Durham University, South Road, Durham DH1 3LE, UK}
\author[0000-0001-9052-9837]{Scott Tompkins}
\affiliation{International Centre for Radio Astronomy Research (ICRAR) and the International Space Centre (ISC), The University of Western Australia, M468, 35 Stirling Highway, Crawley, WA 6009, Australia}
\author[0000-0001-8156-6281]{Rogier Windhorst}
\affiliation{School of Earth and Space Exploration, Arizona State University, Tempe, AZ 85287-1404, USA}
\author[0000-0001-9491-7327]{Simon Driver}
\affiliation{International Centre for Radio Astronomy Research (ICRAR) and the International Space Centre (ISC), The University of Western Australia, M468, 35 Stirling Highway, Crawley, WA 6009, Australia}
\author[0000-0001-6650-2853]{Timothy Carleton}
\affiliation{School of Earth and Space Exploration, Arizona State University, Tempe, AZ 85287-1404, USA}
\author[0000-0003-1625-8009]{Brenda Frye} %%% brendafrye@gmail.com
\affiliation{Department of Astronomy/Steward Observatory, University of Arizona, 933 N Cherry Ave, Tucson, AZ, 85721-0009, USA}
\author[0000-0002-5899-3936]{Leo Fung}
\affiliation{Centre for Extragalactic Astronomy, Department of Physics, Durham University, South Road, Durham DH1 3LE, UK}
\affiliation{Institute for Computational Cosmology (ICC), Department of Physics, Durham University, South Road, Durham DH1 3LE, UK}
\author[0000-0002-3783-4629]{Jiashuo Zhang}
\affiliation{Department of Physics, The University of Hong Kong, Pokfulam Road, Hong Kong}
\author[0000-0003-3329-1337]{Seth H. Cohen} %%% seth.cohen@asu.edu
\affiliation{School of Earth and Space Exploration, Arizona State University, Tempe, AZ 85287-1404, USA}
\author[0000-0003-1949-7638]{Christopher J. Conselice}
\affiliation{Jodrell Bank Centre for Astrophysics, Alan Turing Building, University of Manchester, Oxford Road, Manchester M13 9PL, UK}
\author[0000-0001-9440-8872]{Norman A. Grogin}
\affiliation{Space Telescope Science Institute, 3700 San Martin Drive, Baltimore, MD 21218, USA}
\author[0000-0003-1268-5230]{Rolf A.~Jansen}
\affiliation{School of Earth and Space Exploration, Arizona State University, Tempe, AZ 85287-1404, USA}
\author[0000-0002-6610-2048]{Anton M. Koekemoer}
\affiliation{Space Telescope Science Institute, 3700 San Martin Drive, Baltimore, MD 21218, USA}
\author[0000-0002-6150-833X]{Rafael Ortiz III} %%% rortizii@asu.edu
\affiliation{School of Earth and Space Exploration, Arizona State University, Tempe, AZ 85287-1404, USA}
\author[0000-0003-3382-5941]{Norbert Pirzkal} %%% npirzkal@stsci.edu
\affiliation{Space Telescope Science Institute, 3700 San Martin Drive, Baltimore, MD 21218, USA}
\author[0000-0001-9262-9997]{Christopher N. A. Willmer} %%% cnawillmer@gmail.com
\affiliation{Steward Observatory, University of Arizona, 933 N Cherry Ave, Tucson, AZ, 85721-0009, USA}

%\collaboration{20}{(PEARLS team authors)}



\begin{abstract}
A distinct power-law break is apparent at $m_{\rm{AB}}\sim 21$ in the deep Near-Infrared PEARLS-JWST galaxy counts. The break  becomes more pronounced at longer wavelengths, with 
the counts slope flattening smoothly with apparent magnitude in the shortest band used at  $0.9\,\mu$m, trending towards an increasingly broken slope by the longest wavelength passband of JWST NIRCam, $4.4\,\mu$m. 
This behaviour is remarkably well predicted by the \texttt{GALFORM} semi-analytical model of galaxy formation.
%, established by independent data at other wavelengths and calibrated with local luminosity functions. 
We use the model to diagnose the origin of this behaviour.
We find that the features that are responsible for the break are: 1) the inherent break in the luminosity function; 2) the change in the volume element with redshift; 3) the redshift-dependent nature of the $k$-correction.
We study the contribution to these effects by early and late-type galaxies, using as a proxy for morphology the bulge-to-total stellar mass ratio. 
We find that the way in which ellipticals populate the bright end of the luminosity function while spirals dominate the faint end is preserved in the galaxy number counts, with a characteristic stellar mass at the break of $\sim 10^{10}M_\odot$. 
We also find that the shape of the number counts is mainly driven by galaxies with relatively low redshift ($z\lesssim 2$) for the PEARLS observational limit of $m_{\rm{AB}}\lesssim 28$.
%, centred at a stellar mass of $\simeq 10^{10}M_\odot$ at redshifts 1$<$z$<$3. 
%Over this redshift range the flux decline with increasing luminosity distance is approximately countered by the negative k-correction for early-type galaxies and thus the knee-shaped luminosity function is imprinted on the IR counts. 
%Furthermore, we find consistency with the observed 40\% proportion of spheroidal galaxies morphologies now identified in deep JWST images, spanning cosmic noon and earlier 1$<$z$<$5.
%Deeper predictions by GALFORM test of the limits of our assumptions for galaxy formation and merging evolution, where we predict luminous ($M_{AB}<-21$), bulge dominated galaxies to be absent above z$>$5.
We give a comprehensive description of why the galaxy number counts in the near-infrared PEARLS-JWST observation look the way they do and which population of galaxies is dominant at each apparent magnitude. 
\end{abstract}



\section{Introduction} \label{sec:intro}
The practice of counting objects in the sky with the aim of understanding the structure of the Universe goes back to the pioneering studies of \cite{herschel1785}, \cite{kapteyn1922} and \cite{seares1925}. 
These studies counted stars well before the concept of galaxies was introduced and helped discover the flat and disk shape of our own Milky Way.
With the discovery of galaxies beyond the Milky Way, the focus shifted from stars to galaxies number counts. 
One of the first quantitative works that connected galaxy luminosities to their observed counts was from \cite{hubble1934}. With the same aim \cite{shapley1932} created a catalogue of bright galaxies while \cite{zwicky1937} studied for the first time galaxy number counts within galaxy clusters.
\cite{sandage1961}, for the first time, used galaxy number counts to infer the geometry and the evolution of the Universe. 
This inspired \cite{noonan1971} which made use of deeper observation and incorporated in their analysis also the evolution of galaxy properties. This opened up the debate on whether galaxy counts are more affected by the cosmological framework or galaxy evolution. 
\cite{tinsley1977} and \cite{tinsley1980} proposed ways to include the galaxy evolution modelling into the interpretation of the counts claiming that it was more important than the underlying cosmology. Years later the importance of cosmology was reintroduced by \cite{driver1996} with a study of only elliptical galaxies to minimize the effect of galaxy evolution. 
%Galaxy number counts as a function of apparent magnitude are one of the most fundamental statistical tools that is able to connect extragalactic observational data to their cosmological framework 
The connection between galaxy number counts, galaxy properties and cosmological framework, making use of increasingly deeper galaxy surveys has been the object of many studies \citep{shanks84,jones91,metcalfe91,broadhurst92,yasuda01,cooray16,marr23}.
%have been applied to galaxy surveys to understand the cosmological framework. 
One important step towards modern galaxy counting using CCDs was made by \cite{tyson88}, who pointed out the excess of faint blue galaxies. 
The debate on why we see very high counts of blue galaxies at faint magnitudes continued for years until \cite{glazebrook95} and \cite{driver95} took advantage of the superior resolution of the Hubble Space Telescope (HST) to show that the reason was the rapid evolution of low to intermediate-mass galaxies driven by the decline of the cosmic star formation history (CSFH) trough cosmic time \citep{lilly96,madau_dickinson_14} and the associated downsizing, i.e. the transfer of the star formation activity to smaller scale galaxies \citep{cowie96}. 
Another important contribution was made by \cite{metcalfe95} who collected all the number counts observed through the 90s in the optical $B$ and $r$ bands, to finally reach the infrared $K$ thanks to the observation from \cite{cowie94}. 
The study of the number counts at different wavelengths is crucial as galaxies have completely different behaviours moving from the submillimeter, infrared, optical to ultraviolet. 
\cite{driver94} made one of the first multi-wavelength analyses of the number counts, using four different optical filter measurements with the aim of constraining a model in which dwarf galaxies play a special role in addressing the faint blue galaxy excess aforementioned. 
All the multi-wavelength studies of number counts rely on precise modelling of the $k$-correction to convert the flux received in the observer frame to the flux emitted in the rest frame of the galaxy \citep{coleman80,hogg99,blanton07}. $k$-correction could be very different for different galaxy types as it relies on the associated shape of the spectral energy distribution (SED).
One advantage of the near-infrared wavelengths is that the SED of different galaxy types remains roughly self-similar (see for example fig~1 of \citealt{cowie94}). 
Thanks to the ability of the Wide Field Camera 3 (WFC3) on the Hubble Space Telescope (HST) to have high-resolution imaging and spectroscopy from the ultraviolet to the near-infrared, \cite{windhorst11} revitalised the interest in galaxy counts showing their relation to the extragalactic background light (EBL). Since the EBL can be used to construct cosmic ray attenuation models, \cite{windhorst11} highlighted the importance of galaxy number counts also in practical applications.
%,  and hence their importance in constructing cosmic ray attenuation models. 
The relation between number counts and EBL was initially analysed by \cite{madaupozzetti00} where they showed that for the EBL to converge, the slope of the count has to be smaller than 0.4 which is what happens in every band at fainter magnitudes, while for the brighter part it is approximately 0.6. This implies that the broken power-law in the counts is needed to have a finite EBL and galaxies dominating the break are those contributing the most to the EBL. Intuitively the flattening of the slope of the galaxy number counts can be seen as a balance between the increasing number of faint galaxies and the decreasing of the individual flux contribution due to cosmological dimming. \cite{driver16} further scrutinised the relation between galaxy number counts and EBL showing additional effects related to different galaxy populations and galaxy evolution processes. This suggests that to have a comprehensive view of the processing regulating the shape of the number counts and their relation to the EBL, galaxy formation models are an excellent resource.
%with the advent of PAERLS we decided to use the best number counts using the best galaxy formation model, galform.
%
%Several surveys in different passbands have collected precise measurements for the number counts ranging from the ultraviolet to the submillimeter. 
%Each wavelength can carry different information as it tracks different properties of the galaxy and can be affected by dust extinction by different amounts.

In this context, we decided to use the GALFORM semi-analytical model of galaxy formation \citep{cole:2000,lacey:2016} and test it on the JWST near-infrared number counts collected by the PEARLS team \citep{windhorst23}.
%
%the fully functioning JWST provides unprecedented data in the infrared regime, pushing the observational limit for the infrared number counts up to magnitudes of $m_{\rm{AB}}\sim28$. 
%
The wide filters of the Near Infrared Camera (NIRCam) on JWST can cover the reddest part of the optical spectrum at $0.7 \,\mu$m up to the NIR part at $4.4\,\mu$m, while the Mid-Infrared Instrument covers $4.9-28.8\,\mu$m. 
The PEARLS team in the first JWST cycle collected exquisite NIRCam images of two unbiased fields that they used to provide accurate deep number counts in the window $0.9-4.4\,\mu$m \citep{windhorst23}.
Combining these new observations for the faint part of the number counts with the results from previous brighter surveys such as GAMA \citep{Driver:2011} and DEVILS \citep{davies21}, the knee at about $m_{\rm{AB}}\sim20$ is clearly visible, offering a perfect observational sample for our analysis.
%the counts show a clear broken power-law behaviour with a knee at approximately $m_{\rm{AB}}\sim21$.
This behaviour has been previously predicted by galaxy formation models (see for example \citealt{cowley18} and \citealt{yung22}). In this work, after checking that the model predictions accurately reproduce the observations, we take advantage of the additional information coming from the model to deconstruct the galaxy number counts into the constituent luminosity functions (LF) and put together a comprehensive explanation of how the break in the counts originates and the role that different galaxy populations have at different redshifts.

The motivation of this study is to provide a solid and physically motivated framework of what we can and can not infer from the shape of the number counts. 
This work is timely because with JWST there is a renovated interest for faint high-redshift objects. 
This could lead to the misconception that the faint part of the number counts could be extremely sensitive to the presence or absence of these extremely high redshift faint objects. 
Although we do not completely exclude the possibility that high-redshift objects can slightly affect the very faint end of the number counts, in this study we show that up to the sensitivity of the PEARLS observation ($m_{\rm AB}\lesssim 28$) we can accurately reproduce the shape of the counts by using only luminosity functions below redshift 2. 
This suggests that the faint end of the galaxy number counts on its own is not sufficient to infer the nature of dark matter, because it is still dominated by the luminosity of these low redshift objects.
By giving a fully comprehensive explanation of the factors that shape the number counts, we set the ground for further studies that aim to interpret the physics behind galaxy statistics, in particular by identifying all the observational biases that arise from the change in reference frame in an expanding Universe.

In \S2 we outline the GALFORM semi-analytical model and the PEARLS observations that we will compare our predictions to; in \S3 we present our results starting from our prediction for the galaxy number counts to their interpretation in terms of galaxy populations and the role they play in the origin of the break; finally in \S4 we summarise our findings and conclusions.

\section{Methodology}

\subsection{The GALFORM model}
To make predictions for observable quantities using a physical model we use the GALFORM semi-analytical galaxy formation model \citep{cole:2000}.
GALFORM follows the hierarchical formation and growth of dark matter haloes, the sites of galaxy formation, as well as modelling the main processes regulating the baryonic component and the formation of galaxies. 
Here, we break with the standard approach followed over the past 25 years of using halo merger histories extracted from high resolution $N$-body simulations \citep{Kauffmann1999, Benson:00}, and instead use trees generated using a Monte Carlo approach. The reason for this is to thoroughly investigate the convergence of the counts. The Monte Carlo approach displays considerable flexibility, allowing the count predictions to be tested by varying the mass resolution of the trees and by targetting examples of trees that may be extremely rare within a $N$-body simulation volume.    
%The way the dark matter component is treated in hierarchical models is through the construction of merger trees \citep{lacey93}. Merger trees describe the collapse of dark matter halos due to gravitational instability and follow the growth of dark matter structures through accretion and merging. There are two main ways in which merger trees can be built. The first method consists of making use of only dark matter only N-body simulations. 
The Monte-Carlo technique is based on the Extended Press-Schechter theory to predict the probability of having a progenitor/descendant halo of a specific mass at a specific redshift \cite{lacey93}. This formalism can be exploited to build an algorithm for generating merger histories \cite{cole:2000}. The Monte Carlo approach can be adjusted to improve the agreement with the results of $N$-body simulations (see for example \citealt{parkinson07})
%, but they have different limitations and different advantages, that need to be evaluated depending on the aim of the study.
%In this work, we decided to use the Monte Carlo approach because of its flexibility in the creation of the merger trees.

When using an $N$-body simulation to create a set of merger trees, we are forced to use the intrinsic resolution of the dark matter halos in the specific $N$-body simulation, along with its adopted cosmology.
%This means that once the $N$-body simulation is done, the mass resolution and the cosmology are set.
%and the only changes applicable are those related to the baryonic physics controlling the galaxy formation process. 
%On the other hand, when using the Monte Carlo approach, we could still have changes to the nature of dark matter, for example, we can see how much the resolution of dark matter halos can affect our results or we can explore how different power spectra for different dark matter particles play a role in the formation of the Universe.
The resolution of dark matter halos could be of critical importance when studying the high redshift universe where galaxies form in very small halos, and ignoring them could result in unrealistic redshift distributions. Also, the frequency of the simulation snapshots controls the time resolution of the halo merger histories. Often simulations designed to study galaxy properties at low redshift have a relatively small number of high redshift snapshots. With Monte Carlo merger trees, the number of timesteps can be chosen to ensure convergence in the predicted galaxy properties at high redshift. 
Since in the study of number counts we don't need information regarding the spatial location of galaxies, we are not forced to use the $N$-body approach. 
The spatial location would be needed only if we wish to create a lightcone, as for example in \cite{manzoni24}, or explicitly predict the clustering of the galaxies as a function of separation (rather than predicting their effective clustering bias). 
However, for the study of the number counts the spatial distribution of galaxies is not required, and hence we can take advantage of all the other characteristics of the Monte Carlo approach.
%Whether we decide to create the merger trees through an N-body or a Monte Carlo approach, GALFORM will independently take care of all the processes that control the galaxy formation within the dark matter halos.

The main galaxy formation processes modelled by GALFORM are extensively discussed in \cite{lacey:2016}. We repeat the main ones here for convenience:
\begin{itemize}[itemsep=1mm, parsep=0pt]
    \item effects of the dark matter halo masses and merger history on the formation and evolution of galaxies;
    \item formation of galactic disks through shock heating and radiative cooling of the gas inside the halos;
    \item quiescent star formation in the disk and nuclear starbursts triggered by mergers or disk instabilities;
    \item three different feedback mechanisms that suppress star formation: supernova-driven winds, AGN heating of the hot halo and photoionization of the intergalactic medium;
    %\item formation of spheroids and activation of starbursts by galaxy mergers that are the consequence of the dynamical friction of the intergalactic medium while moving towards the centre of the common dark matter halo. Starbursts and spheroid formation can also be triggered by bar instabilities;
    \item conservation of angular momentum between the three components in which a galaxy is modelled, i.e. disk, bulge and dark matter halo (this allows for the calculation of galaxy sizes);
    \item chemical enrichment of stars and gas. Together with the star formation, the chemical enrichment is used to link the galaxy to a stellar population synthesis model and obtain luminosities; 
    \item attenuation of starlight by dust, modelled using the results of a radiative transfer calculation.
\end{itemize}

These processes are modelled through a set of coupled differential equations with a set of adjustable parameters as listed in table 1 of \cite{lacey:2016}. The search for the best set of values for those parameters is what is usually referred to as the calibration of the model. 
These adjustable parameters are varied within a range of values suggested by theory, and the predictions of the simulation are compared to some observational constraints mainly in the local universe.  The model presented here was calibrated by hand (see e.g. \citealt{Baugh:2019}); a more modern, reproducible approach is described in \cite{elliott21} and \cite{Madar24}.
%The adjustable parameters are classified as primary and secondary depending on the effect that they have on our key observational constraints. 
Observational constraints are classified as primary and secondary, depending on their importance. Once the parameters are set in order to have a good fit of the primary observational constraints, a minor calibration is performed on the secondary observational constraints. 
This minor calibration on the secondary observational constraints has the requirement of not degrading the primary observational constraints.
We list here the primary observational constraints that have been used to find the best values for the parameters:
\begin{itemize}[itemsep=1mm, parsep=0pt]
    \item optical $b_J$ and near-infrared K-band luminosity functions at $z=0$;
    \item observed gas content of galaxies through the HI mass function at $z=0$;
    \item fraction of early vs late-type galaxies\footnote{A proxy for the visual morphology is to use the disk to bulge light ratio or the galaxy's Petrosian concentration index, $c$.} at z=0 as a function of luminosity in the $r$-band;
    \item relation between the mass of the central supermassive black hole and the bulge of the galaxy at $z=0$; 
    \item luminosity function in the rest frame near-infrared $K$-band as a proxy for the stellar mass function at $z=0.5$, $1$ and $3$; 
    \item submillimeter galaxies (SMG) number counts and redshift distribution from deep surveys at $850\mu$m;
    \item far infrared number counts at the wavelengths measured from Hershel and Planck, i.e. $250$,$350$ and $500\mu$m which can probe the population of dusty star-forming galaxies at low and intermediate redshifts;
    \item rest-frame far-UV luminosity function of Lyman Break Galaxies (LBG) at $z=3$ and $6$. 
\end{itemize}
The secondary observational constraints, that we remind are used to refine the model only if they do not degrade the quality of the primary constraints are:
\begin{itemize}[itemsep=1mm, parsep=0pt]
    \item Tully-Fisher relation between the luminosity and the circular velocity of spiral galaxies at $z=0$;
    \item relation between galaxy size (half-light radius) and luminosity in the $r$-band (as a proxy for stellar mass) at $z=0$, for early and late-type galaxies;
    \item stellar metallicities vs luminosity relation for early-type galaxies at $z=0$.
\end{itemize}
The key point to understanding the role of these constraints is that they are not used to define the evolution of the model with redshift. Instead, they are used once to find the best value for the set of free parameters of the differential equations that govern all the physical processes that drive galaxy evolution. 
To give a practical example, the way the supernovae feedback is modelled is through the rate of mass of gas ejected from the galaxy, and this is assumed to be proportional to the instantaneous star formation rate $\psi$:
\begin{equation}
    \dot M_{\rm{ejected}}=\left( \frac{V_C}{V_{\rm{SN}}}\right)^{-\gamma_{\rm{SN}}}\psi
\end{equation}
where $V_C$ is the circular velocity of the galaxy, $V_{\rm{SN}}$ and $\gamma_{\rm{SN}}$ are adjustable parameters that we aim to set. These parameters are allowed to vary over a range loosely suggested by the physical process (see, for example, the table in \citealt{elliott21}).  The observational constraints are purely used to find the best value SNe feedback parameters.
%$\gamma_{\rm{SN}}$ (and of all the other parameters in the equation described in \citealt{lacey:2016}).
Once these parameters are fixed to their best values
%, those observational constraints are not used in any way and 
the whole predictive power of galform is stored in the set of coupled differential equations used.

\subsection{The PEARLS observations}
The Prime Extragalactic Areas for Reionization and Lensing Science (PEARLS) is a scientific project led by R. Windhorst that takes advantage of the 110 hours of guaranteed time on JWST to study the very early universe and in particular the assembly of galaxies with unprecedented precision \citep{windhorst23}. 
With this aim, they selected the following targets\footnote{We list here all the PEARLS targets, although not all of them are suitable for number counts, so not all of them are used in this work.}:
\begin{itemize}[itemsep=1mm, parsep=0pt]
    \item 3 blank fields;
    \item 7 galaxy clusters that exhibit exquisite features of strong gravitational lensing;
    \item 2 high-redshift proto-clusters;
    \item 2 high-redshift quasars; 
    \item 1 low-redshift spiral galaxy enlightened by a close elliptical.
\end{itemize}
For all of these targets, they performed photometry in seven NIRCam filters covering the window $0.9-4.4\,\mu$m. The NIRCam instrument allows for the imaging of two areas of sky slightly offset (two squares of $2.2' \times 2.2'$ separated by $0.7'$). 
When observing galaxy clusters, this feature makes it possible to observe simultaneously an area of the sky just outside the cluster, generally referred to as the `parallel' or the `non-cluster' field (or module). 
The results on the galaxy number counts summarised in the overview paper \citep{windhorst23} come from the analysis of one of the three blank fields and a non-cluster module. 
In particular, the blank field is the JWST IRAC Dark Field (JWIDF) while the non-cluster module is the parallel field next to the El Gordo cluster at $z=0.870$.  
The JWIDF and the El Gordo non-cluster module cover respectively (at the time of their first epoch observation) $61.2'' \times 109.3''$ and $150''\times 150''$ (see figs.~2 and 3 of \citealt{windhorst23}).
To create the catalogue used for the number counts, they ran SourceExtractor \citep{sextractor} and ProFound \citep{profound} on the NIRCam reduced images.
In the case of JWIDF all the 10 NIRCam detectors have been used\footnote{NIRCam is made of two modules, A and B, each of them having 4 short wavelength detectors ($0.6-2.3\, \mu\rm{m}$) and 1 long wavelength detector ($2.3-5.2\, \mu\rm{m}$).}. For the El Gordo parallel field, they only used 5 detectors (as the others were used for the cluster).
To make sure that the El Gordo parallel field is not affected by the presence of the nearby cluster Windhorst et~al. checked the agreement in the number counts with the filters in common with JWIDF.
The star-galaxy separation procedure that has been used to clean stars from the catalogue in both fields is based on the FWHM vs apparent magnitude plot. The algorithm is described in \cite{windhorst11}.
To complement the deep and faint NIRCam counts, \cite{windhorst23} collected data from brighter surveys. These include GAMA \citep{Driver:2011} and DEVILS \citep{davies21} and for the shorter wavelengths filters, they also used the deepest HST counts available. For a complete description of the complementary data used here, see \cite{koushan21}.
Here, all the data used for the number counts are shown in Fig.~\ref{fig:counts} as grey dots and their relative fit as a blue line (compare the original data from figure~9 and 10 in \citealt{windhorst23}).
%instead of considering the individual data points for the number counts in \cite{windhorst23}, we decided to use the fit that they obtained (as shown by the black line in their figure~9 and 10). 
Note that the fit extends brightwards and faintwards of the magnitude range actually sampled by the data.

\section{Results}
\subsection{Testing the model prediction}
\begin{figure*}
\plottwo{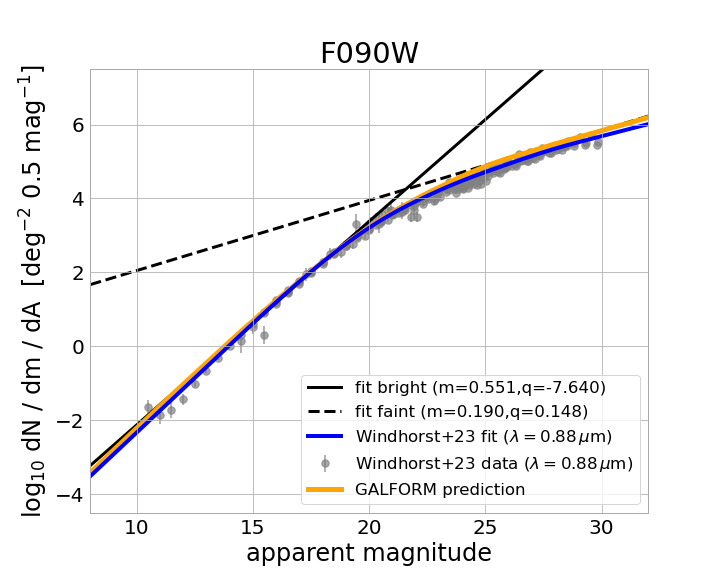}{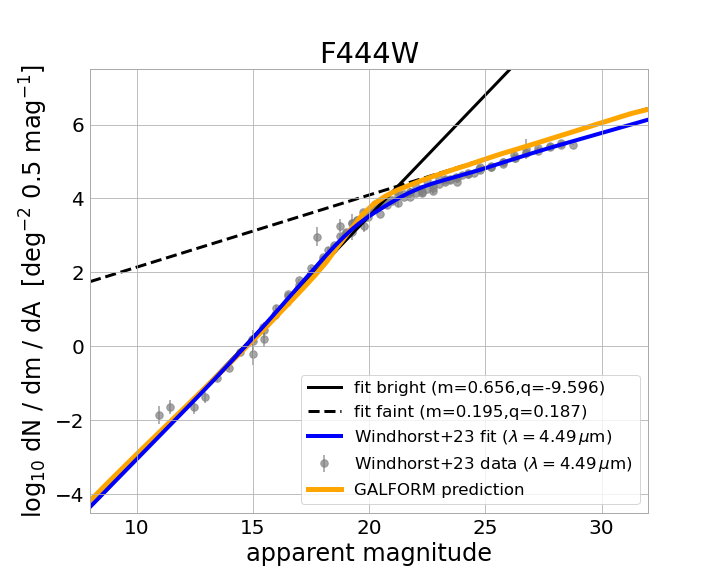}
\caption{Number counts predicted using GALFORM (orange lines) together with the data collected in \cite{windhorst23} (grey dots) and the respective double power-law fits (blue lines). In the left panel, we plot the counts for the shortest wavelength filter used by \citealt{windhorst23}, which is the NIRCam F090W, while in the right panel, we show the longest wavelength counts, F444W. The break in the counts' slope is more pronounced at longer wavelengths. The black lines are power-law fits to the GALFORM counts when considering separately the results on the left and right sides of the break. The intersection of these two lines can be used to precisely locate the break or change in the slope of the counts.
\label{fig:counts}}
\end{figure*}

We now test the model against the observations collected by the PEARLS team in \cite{windhorst23}. In their fig.~9 and 10, they show fits to the galaxy number counts for the JWST NIRCam filters in the range $0.9 - 4.4\, \mu$m and other brighter surveys both from space and ground-based. %We decided to use their power-law fit to compare their observed number counts to the counts we simulated using the GALFORM model. 
We plot the observations from \cite{windhorst23} with grey dots and their fits with a blue line in Fig.~\ref{fig:counts} where we compare it to the simulated galaxy counts in two representative filters (orange lines). F090W (left panel) is the shortest wavelength filter used by the PEARLS team centred at about $0.90\,\mu$m while F444W (right panel) is the longest wavelength filter, centred on about $4.44\,\mu$m. 
%
%The orange line shows the simulated galaxy number counts for the same filters.
%, with the error bars showing the Poisson noise arising from the number of objects simulated. Note that these errors are not connected to the field size of the JWST observations, but to the number of realisations generated for each halo merger history. 
We performed a separate power-law fit to the simulated counts for the bright (apparent magnitude between 10 and 20, black solid line) and faint (apparent magnitude between 25 and 30, black dashed line) parts of the counts, which tend to power laws. We will refer to the intersection between the two linear fits as the knee of the counts. 
We report the parameters of the fits in the legend of Fig.~\ref{fig:counts}. It is interesting to note that the slope of the power-law in the faint regime is very similar at both wavelengths, with $0.190$ for filter F090 and $0.195$ for filter F444W. 
A bigger difference between the two filters is seen, however, at brighter apparent magnitudes, with a slope of $0.551$ found for the F090W counts compared with $0.656$ for F444W, making the knee more pronounced at longer wavelengths. 
We will give a full explanation of this feature in Sect.~\ref{sec:interpretation} and devote the full Appendix~\ref{app} for the technical details. 
However, we anticipate that this effect is due to the combination of the band-shifting and cosmological dimming due to the increasing redshift, when this combines with the shape of the spectral energy distribution (SED) of the objects considered. 
Our model can reproduce the observed counts quite accurately with a small offset in the faint end of the counts in F444W. Most importantly, they are impressively accurate in predicting the slope of the double power law in all the filters, implying that the physics responsible for the knee is included in the GALFORM models. 
This also excludes the possibility of blaming new physics for this clear feature in the galaxy number counts. 
Although we chose to include only the counts for these two representative filters we note that the GALFORM model can reproduce with the same level of agreement the counts observed in the other JWST NIRCam filters used in PEARLS, and we reported them in Fig.~Set~1 available in the online version. 
The GALFORM counts from \cite{cowley18} have been reported to enjoy varying levels of agreement with observations made in other bands observed with the JWST (e.g. \citealt{Ling2022,Yang2023,Wu2023, Sajkov2024,Stone2024}).

\subsection{Interpretation of the break in the counts}
\label{sec:interpretation}
One of the key advantages of using a galaxy formation model to interpret the shape of the number counts is that we can deconstruct the counts into the sum of different luminosity functions (LF) at the specific redshifts corresponding to the outputs of the simulation. 
This could be extremely complicated to do with observational data only. The first reason is that we would need precise spectroscopic redshift measurements. This would defeat the purpose of the number counts as a simple and immediate test. Secondly, different redshift samples would be affected by incompleteness in different ways. Moreover, instead of precise redshift outputs, we would need to deal with narrow redshift bins which would affect the purity of the sample.
%This can be done with observation only with the assumption of having precise spectroscopic redshift measurements and by taking sufficiently small redshift intervals to avoid the contamination of galaxies from different cosmological distances. 
Another advantage is that in simulations we can select galaxies with different properties to see how they contribute to the number counts. 
%This is what we have done and the results are presented in this section.

\begin{figure*}
\plottwo{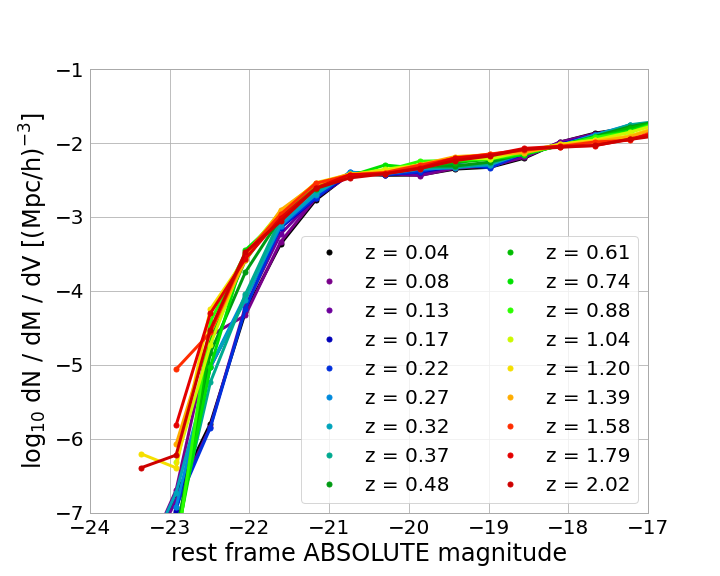}{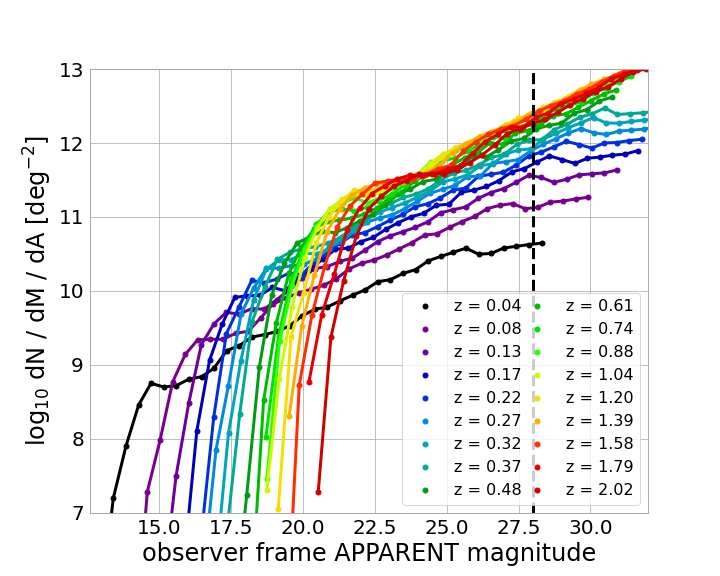}
\caption{Luminosity functions predicted by GALFORM at closely spaced redshifts in the range $0<z<2$, as shown by the key, for the long wavelength filter F444W. Left: we plot the luminosity functions in the traditional rest frame absolute magnitudes vs number of objects per unit magnitude and unit volume. Right: we plot the luminosity function in the units of the number counts, i.e. {\it observer} frame apparent magnitudes vs number of objects per unit magnitude and unit area. The number counts are derived from the integration of the luminosity functions in these units over the redshift range covered by the apparent magnitude limit (for PEARLS observations, this limit is approximately at F444W $\sim28$). It can be seen how the effect of cosmological dimming and $k$-correction moves the luminosity functions horizontally while the change in volume for a fixed area moves the luminosity function vertically. The combination of these effects leads to the break we see in the counts. To compare this effect in all of the NirCAM wide filters, see Figure Set~2 available in the online version.
\label{fig:LF}}
\end{figure*}

\begin{figure}
\centering
\scalebox{1.2}{\plotone{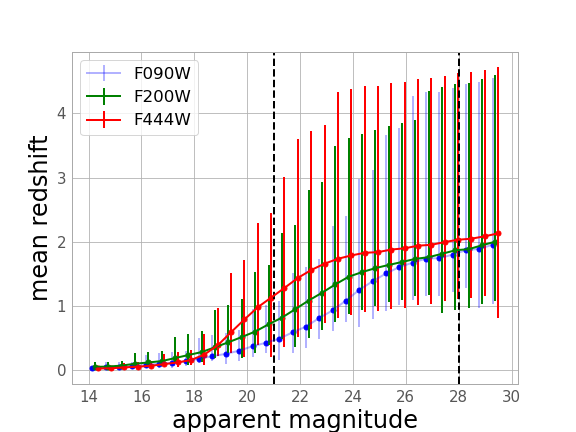}}
\caption{Mean redshift as a function of the apparent magnitude for the objects simulated in three representative NIRCam filters, F090W (blue line), F200W (green line) and F444W (red line). Error bars are the 25th and 75th percentiles of the associated redshift distribution for the related apparent magnitude. The two vertical dashed lines indicate approximately the location of the break in the counts at an apparent magnitude of 21 about and the PEARLS magnitude limit at about apparent magnitude 28. The objects responsible for the shape of the counts lay at relatively low redshifts as indicated by the solid lines ($z\leq2$).
\label{fig:redshift_filters}}
\end{figure}

In Fig.~\ref{fig:LF}, we can see the number counts broken down into their constituent luminosity functions. We have tested that we are able to accurately reproduce the shape of the observed number counts by using only LFs with $z\lesssim 2$, even though the maximum output redshift from our calculations is much higher.  
This is also shown by Fig.~\ref{fig:redshift_filters} where we plot the mean redshift as a function of apparent magnitude for the number counts in three representative filters. 
Up to the PEARLS apparent magnitude limit of $m_{\rm{AB}}\sim28$ (indicated by a vertical dashed line) the mean redshift is below $z\lesssim2$ for all the filters showing that the shape of the number counts is mainly driven by low redshift objects. 
The left panel of Fig.~\ref{fig:LF}, shows the traditional way of plotting luminosity function at different redshifts, i.e. using rest frame absolute magnitudes for the $x$-axis and the number of objects per unit magnitude and unit volume for the $y$-axis. Using rest frame absolute magnitudes means that we are looking at the intrinsic luminosity of the galaxy at the emitted wavelength, ignoring the band shifting and the cosmological dimming due to redshift. This way we can look at the intrinsic properties of the galaxies without worrying about how redshift affects observation. 
%In fact using absolute magnitude is like looking at all the objects as if they were all at a distance of $10$ pc. 
In this traditional way of plotting the luminosity function, we can see the intrinsic galaxy evolution, for example, we see that at the bright end galaxies were a bit brighter at $z\sim2$ than at $z\sim0$ in agreement with the drop in star formation rate density after cosmic noon \citep{madau_dickinson_14}. 
The study of traditional LFs can be very informative for trying to understand galaxy evolution. However, here we are trying instead to understand the galaxy number counts, which are plotted in terms of apparent magnitudes. 
%By using apparent magnitudes instead of rest frame magnitudes, we avoid the need to model the form of the spectral energy distribution of the galaxy and derive a $k$-correction (see \citealt{manzoni24} who also used observed magnitudes for this reason). 
%Rest frame absolute magnitudes instead are not immediately available from observations and can be obtained only through precise modelling of the theoretical spectrum of the galaxy and hence of the $k$-correction to be applied to correct for the band shifting due to redshift. 
%Although this is a process that can certainly be done, it does introduce some model dependency as by using stellar population synthesis modelling, a certain initial mass function and a parametric star formation history have to be assumed. 
%This is why using apparent magnitudes makes the number counts so suitable for testing galaxy surveys. 
Apparent magnitudes are immediately available from observations, are model-independent and do not require a redshift measurement. This makes the use of number counts extremely straightforward as all that needs to be done is plot all galaxies (independently of the redshift) as a function of apparent magnitude. 
On the other hand, rest frame absolute magnitudes are immediately available from simulations and here we will use them, as a tool to understand how number counts are connected to their constituents LFs.
On the right panel of Fig.~\ref{fig:LF} we plot the luminosity functions in an unconventional way which represents the intermediate step to convert a set of standard luminosity functions (left panel of Fig.~\ref{fig:LF}) into number counts (Fig.~\ref{fig:counts}). 
The conversion of the $x$-axis, when converting the LFs from the traditional view in the left panel to the unconventional way in the right panel of Fig.~\ref{fig:LF}, is from rest frame absolute magnitudes into observed apparent magnitude. 
This conversion shifts the luminosity functions due to two main effects. The first effect is the band shifting due to the redshifting of the photons. Emitted photons with frequency $\nu$ at redshift $z$ get a lower frequency $\nu_o$ when reaching the observer frame, the Earth, by $\nu_o = \nu / (1+z)$. 
If the spectra of galaxies were completely flat, the flux at $\nu_o$ and the flux at $\nu$ would be the same. 
%This is trivial in simulation when we have knowledge on the shape of the entire spectrum of the galaxy. 
Since this is not the case, to model this effect we need to know the shape of the spectrum to know the value of the flux both at $\nu_o$ and at $\nu$, so that we can correct the flux by their ratio. We refer to this as the $k$-correction. In other words, whether we have observed fluxes and we want to go rest-frame or we have rest-frame fluxes and we want to go observer-frame, we can do that by adding or subtracting the $k$-correction respectively.
%However, it implies some model dependent correction when doing this with observation as the fact that we are measuring a flux from a filter that has a wavelength different from the original wavelength of the photons means that we need to assume the shape of the spectrum based on the type of galaxy we are observing and enhance or decrease the flux based on the model SED. This process is called $k$-correction. 
The second effect is the decrease in flux as objects move farther away. 
The flux, $F$,  decreases as the square of the luminosity distance: $F_{\nu} = L_{\nu}/(4\pi d_L^2)$, where the luminosity $L_{\nu}$ is simply the flux that the object would have if it were at a distance of $10$ pc. 
%These are the two effects that move a luminosity function left or right when plotting it in the observer frame's apparent magnitudes (as in the right panel of Fig.~\ref{fig:LF}). 
The dimming effect moves the LFs towards fainter apparent magnitudes (moving to the right) while the band shifting can increase (moving to the left) or decrease (moving to the right) the flux based on the shape of the SED. Hence, while the dimming effect applies rigidly to all LF in every filter, the band-shifting can have a different effect on the LF based on the filter being used.
This is because at different redshifts we are looking at different parts of the SED and this can affect how the LFs move in different filters when looking at different redshifts. 

The conversion for the $y$-axis to connect the LF to the number counts, needs to take into account the change in number of galaxies when moving from a simulated box to an area of the sky.  
When considering the classic luminosity function, we plot the number of objects per unit comoving volume per magnitude bin. However, when we compute the number counts we have the number of objects per unit area. This means that we need to take into account the fact that for a fixed area on the sky, we have a volume element which itself is a function of redshift.  
The change in the volume element depends on the underlying cosmology, defined by the components of the energy density of the Universe i.e. matter, $\Omega_m$ (including both baryons and dark matter), and dark energy, $\Omega_{\Lambda}$\footnote{Usually the energy density of radiation, $\Omega_{\gamma}$, and curvature $\Omega_k$, can be neglected.}. 
Assuming the standard values for the cosmological parameters as measured by Plank \citep{plank14}, up to about $z\sim2.5$ the comoving volume increases with redshift. This means that the number of objects gets boosted with increasing redshift when the LFs are considered per unit area rather than per unit volume.
In principle, the LFs plotted this way would be shifted down in amplitude for redshifts higher than $z\sim2.5$ (as the element of comoving volume starts to decrease). However, in practice, this does not affect the number counts, as we can see from Fig.~\ref{fig:redshift_filters}, that the objects dominating the counts come from LF with redshifts lower than $z\sim2$.
When translating the number of objects per unit volume into the number of objects per unit area, we obtain the luminosity functions with the same $y$-axis units as the number counts. 
This is what we get in the right panel of Fig.~\ref{fig:LF}. 
The additional step that is needed to convert the set of LFs in these new units into number counts is an integration over redshift. 
This is exactly what has been done to obtain the GALFORM number counts in Fig.~\ref{fig:counts}.

Given the fact that the integration over all the LFs in the right panel of Fig.~\ref{fig:LF} gives the number counts, we can consider the highest curve as the one making the dominant contribution. 
Moreover, since the outputs of the simulation are enough to sample very well the redshift range studied, we can imagine a line connecting the highest point of the rescaled LF at every apparent magnitude and see how this shows the same break at around magnitude $m_{\rm{AB}}\sim21$. 
We decided to plot the LF curves only the filter F444W as this is the one with the most pronounced break but the simulation accurately reproduces a softer break for the other shorter wavelength filters as is seen in the PEARLS observations (see Fig.~Set~2 in the online version to compare the consistent luminosity function in the units of the number counts for all the filters). 
This can be explained by what we can call a \textit{compression} of the luminosity functions. By this we mean that the LFs are spread over a certain range of apparent magnitudes. 
This range is wider for a shorter wavelength filter, like F090W, and narrower for a longer wavelength filter like F444W. 
By looking at the intermediate filters as well, this effect is readily apparent: going from 0.9 $\mu$m to about $2\mu$m, the LFs get brighter for all the redshift outputs plotted in Fig.~\ref{fig:LF} (preserving approximately the range of apparent magnitude covered). 
However between filters F200W and F277W, some low redshift luminosity functions start to get fainter while the higher redshift LFs keep getting brighter and this effect gets stronger going from filter F277W to F444, creating the \textit{compression effect} just mentioned, which diminishes the range of apparent magnitudes covered (compare the figures in the Fig.~Set~2, available in the online version). 
This is the combined effect of the band-shifting and dimming created by redshift.
In fact, while the dimming effect always decreases the flux with increasing redshift, the band-shifting can increase or decrease the flux depending on whether we are \textit{climbing} or \textit{descending} the SED (see more in Sect.~\ref{sec:SED} and Appendix~\ref{app}). 
Whether the dimming effect and the band-shifting are contributing together in decreasing the flux or contrasting each other at a specific redshift will result in the LF getting brighter or fainter, leading to the \textit{compression} effect and hence a stronger break.
From Fig.~\ref{fig:redshift_filters} it is also clear that different filters track different redshifts exacerbating the effect just explained. 
It is also interesting that the filter F444W is changing quite rapidly redshift around magnitude $\sim20-21$ which corresponds to where the location of the knee in the counts is.

\subsection{Which galaxies dominate the number counts?}
To test whether the break in the counts corresponds to a change in the underlying population of galaxies tracked by a specific filter we decided to look at how different morphologies contribute to the LFs and hence the galaxy number counts. 
We take a simple approach of using the bulge to total stellar mass ratio, denoted by $B/T$,  as a proxy for morphology.
We put the threshold at $B/T=0.5$ so that we can include all the galaxies and consider galaxies with $B/T<0.5$ as \textit{disk-dominated} and those $B/T\geq0.5$ as \textit{bulge-dominated}. 

%We exclude galaxies with exactly $B/T=0$ and $B/T=1$ as they arguably may not have a well-defined morphology but could be in some merger stage. 
%Without going into detail, this is just due to the fact that when galaxies are formed they come with B/T=0 but during mergers, galaxies change their structure to purely bulge ($B/T=1$).

\begin{figure}
\centering
\scalebox{1.2}{\plotone{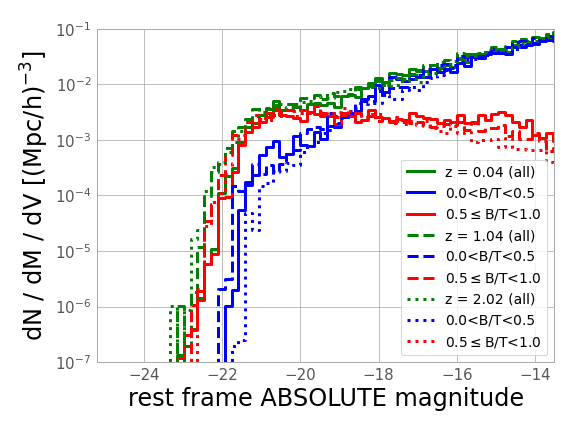}}
\caption{The predicted luminosity function in filter F444W at three redshift outputs as indicated by the key. The different colours indicate different morphologies as classified by the $B/T$ criterion (see text). The similarities of the three LFs plotted indicate that the intrinsic properties of galaxies vary less than the cosmological evolution of observed properties due to redshift (i.e. the LFs in the left panel of Fig.~\ref{fig:LF}, plotted in terms of rest frame magnitude vary less than those in the right panel, plotted against observed magnitude). The contribution of high and low $B/T$ to the LF is the same at every redshift. This means elliptical (bulge-dominated) galaxies dominate the bright part of the LF and spiral (disk-dominated) galaxies contribute to the faint part of the LF.  
\label{fig:LF_BT}}
\end{figure}
\begin{figure}
\centering
\scalebox{1.2}{\plotone{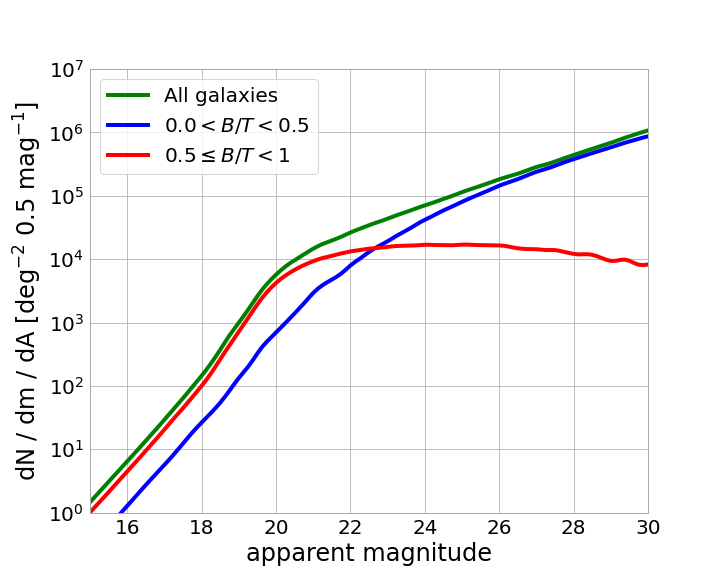}}
\caption{Simulated galaxy number counts in the NIRCam filter F444W. Different line colours are used to indicate morphology through the $B/T$ proxy, as shown by the legend.  
The green line is for all objects. The blue line is for disk-dominated galaxies with $B/T\leq0.5$. The red line shows bulge-dominated galaxies with $B/T\geq0.5$. It is clear that \textit{bulge dominated} galaxies make up the bright part of the counts while \textit{disky} galaxies dominate the faint counts. To fully understand the behaviour of the counts, this figure can be compared with the redshift distributions of the same objects in Fig.~\ref{fig:n_of_z_BT}. 
\label{fig:counts_BT}}
\end{figure}

Fig.~\ref{fig:LF_BT} shows the luminosity function in three representative outputs of the simulation, i.e. $z=0.04$, $z=1.04$ and $z=2.02.$, for the NIRCam filter F444W.
As suggested by the left panel of Fig.~\ref{fig:LF}, the rest frame standard LFs do not change much within the redshift range under consideration ($0<z<2$) which is the redshift range of the galaxies dominating the number counts (see Fig.~\ref{fig:redshift_filters}). 
Fig.~\ref{fig:LF_BT} confirm the self-similar behaviour of the LFs also in terms of populations.
%We use the $B/T$ ratio as a first-order approximation or proxy for a visual morphology, as is common practice in semi-analytical models. 
It is clear how early-type galaxies (high B/T ratios) dominate the bright part of the luminosity function while late-type galaxies (low B/T ratios) dominate the faint part. 
Thus we can loosely treat the single LF at $z=0.04$ as being representative of all the others. 
Whilst doing this, we do not want to ignore the effect of the evolution of intrinsic galaxy properties but we acknowledge that this effect is negligible in shaping the number counts when compared to the cosmological evolution due to redshift. 
This is in agreement with \cite{marr23}.
%With this in mind, we have split the LF into two components: disk-dominated galaxies ($B/T<0.5$) and bulge-dominated galaxies ($B/T\geq0.5$). 
%We use the B/T criterion as a proxy for identifying elliptical early-type galaxies and spiral late-type galaxies. 
%Although here we show only a low redshift luminosity function, we checked that this behaviour holds at higher redshifts. 
The interesting feature is that the transition between early and late type in dominating the number of galaxies of the luminosity function is approximately where the break occurs between the two power-laws. 
We note that in Fig.~\ref{fig:LF_BT} we are only looking at three single luminosity functions and how this converts into number counts is not trivial.
In fact, we need to integrate over the luminosity functions at all redshifts to obtain the number counts. 
We can do this separately for the two populations and this results in Fig.~\ref{fig:counts_BT}.
The green line in Fig.~\ref{fig:counts_BT}, corresponds to the \galform prediction as plotted in Fig.~\ref{fig:counts}, but now we can see the contribution of the two populations of galaxies. i.e. bulge-dominated (red line) and disk-dominated (blue line). How the integration of the single luminosity functions brings to this shape for the number counts can be seen on the right panel of Fig.~\ref{fig:LF}. The combination of the change in comoving volume for a fixed area of the sky and the dimming and band-shifting of the objects due to redshift, give the shape to the counts with the well-defined break in power-law slope. The additional information we find in Fig.~\ref{fig:counts_BT} is how the different contributions of bulge and disk galaxies can describe this broken power-law. In fact, the apparent magnitude of the break corresponds to the change in the dominant population with bulge galaxies dominating the bright part and disk galaxies dominating the faint end. Although this is a simple picture there is an underlying complexity which is related to which redshifts contribute the most at every apparent magnitude. In fact, we stress that number counts include objects at all redshifts. To investigate this further, we drew the redshift distribution for two representative apparent magnitude limits (i.e. considering all the objects brighter than the stated limit). 
\begin{figure*}%[h]
\plottwo{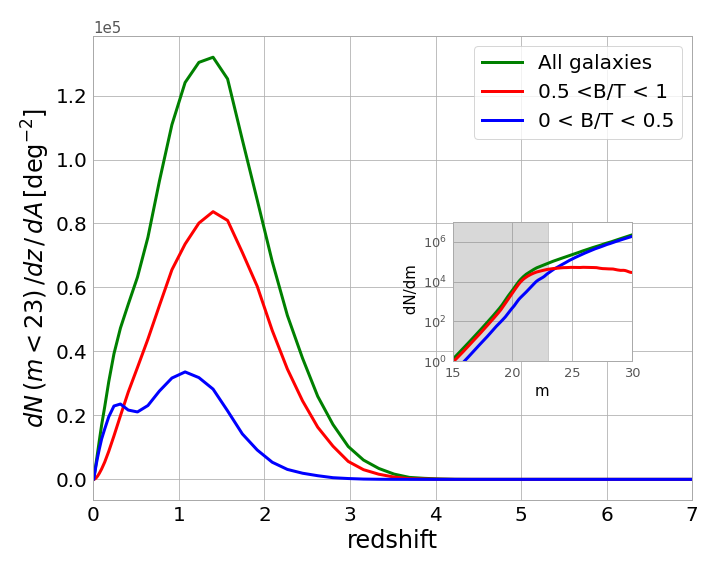}{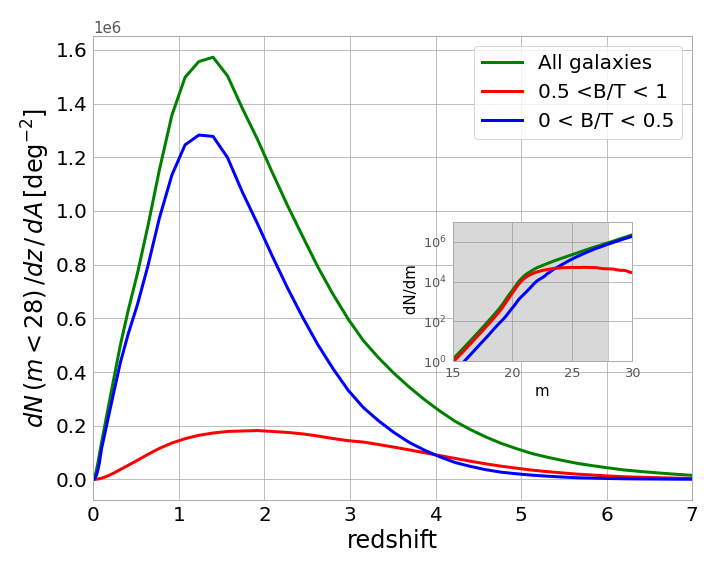}
\caption{Redshift distributions for the objects plotted in Fig~\ref{fig:counts_BT}. Different colours represent different morphologies according to the $B/T$ criterion as in Fig~\ref{fig:counts_BT} (see text for a full description). The two panels show the distributions for different magnitude limits. Left: F444W$<23$. Right: F444W$<28$ (this corresponds to the PEARLS magnitude limit). Bulge-dominated objects dominate the bright counts at most redshifts while the disk-dominated objects dominate the faint counts. Note that the number of galaxies satisfying the brighter limit (left panel) is about one order of magnitude lower than the number of galaxies satisfying the fainter limit (right panel). The shaded area in the insets helps identify which part of the number counts the galaxies in the redshift distribution belong to.
\label{fig:n_of_z_BT}}
\end{figure*}
In Fig.~\ref{fig:n_of_z_BT}, we plot the redshift distribution for a bright magnitude limit of 23 and a fainter magnitude limit of 28 (which corresponds to the PEARLS sample limit). The first thing to notice is how in both panels the peak of the redshift distribution is limited to relatively low redshifts in the range $1<z<2$. The second immediately evident fact is that when considering the bright limit (left panel), bulge objects are dominating in number while the overall trend is inverted when considering the faint limit (right panel). Another more subtle feature is that the peak in the $n(z)$ of the bulge galaxies is always slightly higher in redshift than for disk galaxies. This is in agreement with the fact that elliptical galaxies have bigger stellar masses than disk galaxies and hence intrinsically brighter. This makes them still visible at higher redshift when disk galaxies would be already too faint to be seen. 

\begin{figure*}%[h]
\plottwo{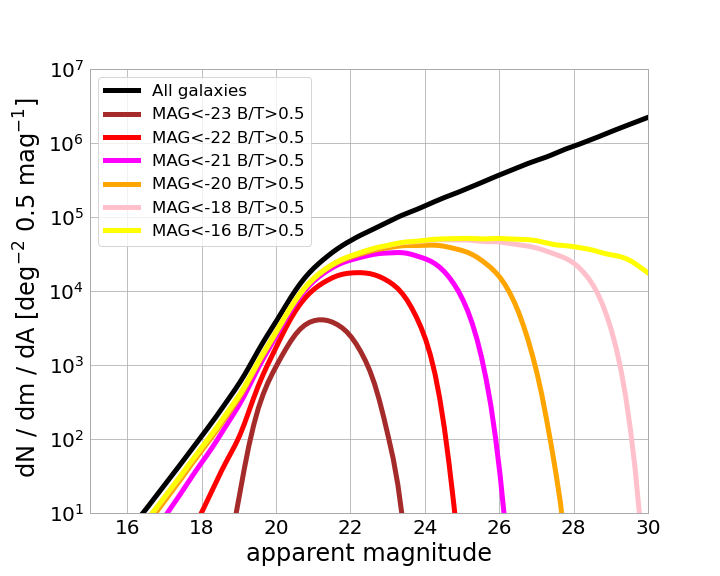}{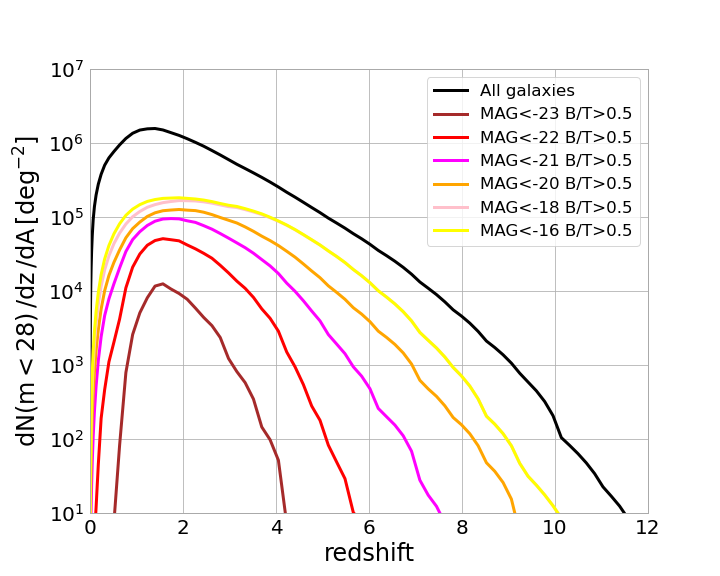}
\caption{Left: simulated galaxy number counts in the filter F444W, built by using luminosity functions of objects satisfying different criteria indicated by the line colour: black, all objects, all the other lines are for objects with a bulge-total mass ratio greater than 0.5 ($B/T\geq 0.5$), selecting bulge-dominated galaxies with the aim of considering only early-type galaxies. Apart from the black line, all the other lines only use objects with a rest-frame absolute magnitude brighter than the threshold indicated in the legend. Right: redshift distribution in logarithmic units (on the $y$-axis) for the same objects in the left plot when applying an apparent magnitude limit of F444W$<28$.
\label{fig:counts_absmag}}
\end{figure*}

\subsection{Contribution of elliptical galaxies }
We focus now on the contribution of elliptical galaxies to the counts. The reason for this is that elliptical galaxies are to an extent less sensitive to galaxy evolution, prioritising the cosmological effect as suggested in \cite{driver1996}.
As in the above analysis, we rely on the bulge to total stellar mass as a proxy for the morphology of the galaxy and consider ellipticals as galaxies that have a ratio $B/T\geq0.5$. 
For these galaxies, we consider the number counts after applying a selection on their rest frame absolute magnitude. 
The reason for that can be seen in the left panel of Fig.~\ref{fig:LF}. 
When looking at the luminosity functions in the rest frame (left panel of Fig.~\ref{fig:LF}), we can assume that approximately they all have the same shape in the redshift range of interest. 
There is little variation in the bright end, due to galaxy evolution (presumably the activity of AGN feedback on massive galaxies) leading to less change in the LF than when plotting the LF against the observed magnitude.  
Moreover, we have noted that all the LFs in the redshift range of interest have the same behaviour when split by morphology, i.e. bulge galaxies dominate the bright end and disk galaxies dominate the faint end (see Fig.~\ref{fig:LF_BT}). 
This implies that if we select galaxies brighter than a given rest frame absolute magnitude, and vary this selection, then we are progressively sampling a different galaxy population.
When the rest frame absolute magnitude limit is bright the population is mainly made up of ellipticals. 
When the limit is at fainter rest frame absolute magnitudes instead, the population includes both bulge and disk-dominated galaxies. 

%To study how elliptical galaxies of different intrinsic luminosities affect the counts, we selected only elliptical galaxies using the $B/T$ criterion and then we cut in rest frame absolute magnitude as described above. 
In the left panel of Fig.~\ref{fig:counts_absmag} we show the counts for six representative rest frame absolute cuts between $-23$ (probing the bright end) and $-16$ (deep into the faint end). 
The part of the luminosity function these limits are tracking can be seen by examining Fig.~\ref{fig:LF_BT}. 
%On the right panel, we can see their relative redshift distribution on a logarithmic scale. 
If we look for example at the brighter limit of $M<-23$ we are selecting only the very bright part of the luminosity function where the high luminosity ellipticals are the most numerous (brown line in Fig.~\ref{fig:counts_absmag}). 
Looking at the left panel of Fig.~\ref{fig:counts_absmag}, we note that these very bright objects are not responsible for the bright apparent magnitudes in the number counts but they are mostly located at the position of the break in the counts. If we allow the rest frame absolute magnitude limit to sample slightly fainter objects, we can see that we need those objects to get closer to the total counts (the black line). 
For example, with a limit of $M<-21$ (magenta line) we are getting close to the total number of objects in the counts. A rest frame absolute magnitude limit of $M<-21$ corresponds roughly to the break in the rest frame luminosity function as shown in Fig.~\ref{fig:LF_BT}. 
This means that we need all the objects brighter than the break of the luminosity function to reproduce the break in the number counts. For fainter magnitudes than the break in the number counts we really need the disk galaxies to recover the overall shape of the counts. 
By using only ellipticals, we note that by taking galaxies with increasingly fainter rest frame absolute magnitude limits we can see the cutoff in the number counts moving to fainter apparent magnitudes.
In the right panel of Fig.~\ref{fig:counts_absmag}, we can see the relative redshift distributions on a logarithmic scale, for an apparent magnitude limit of 28. 
We have already seen that for the total counts, we expect the mean of the redshift distribution to be below $z\sim2$. 
Here we can see what the whole redshift distribution for ellipticals looks like and the difference when we include only intrinsically luminous objects (as with the brown line) and all the galaxies up to a faint limit (as shown by the yellow line). Given the field of view of NIRCam, we expect to have one object per field of view when the $y$-axis is about $7\times10^2 \frac{\rm{galaxies}}{\rm{deg}^2}$. %743.6 galaxies per squared degree (NIRCam FoV)
Hence when compared to the observations from NIRCam, it makes sense to only consider the counts above that threshold. 
It is interesting to note that when we consider objects more luminous than absolute magnitude $-21$, we are almost considering all the ellipticals that have an effect in shaping the number counts. 
%we can see that the redshift distributions don't change much by considering fainter limits. 
In fact, the number counts already look almost complete at the bright end when using that magnitude limit.
\begin{figure*}%[h]
\plottwo{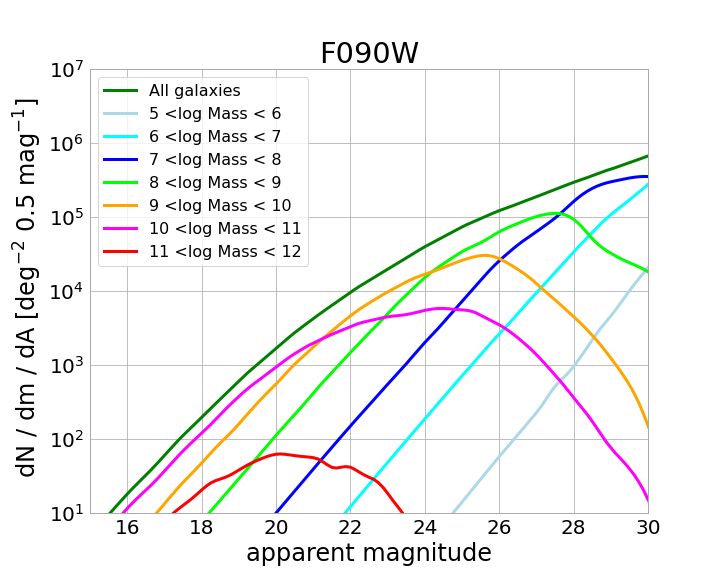}{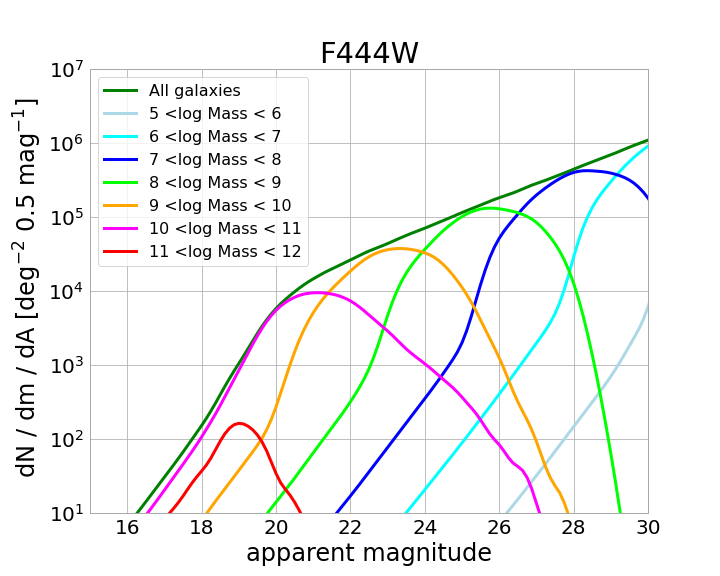}
\plottwo{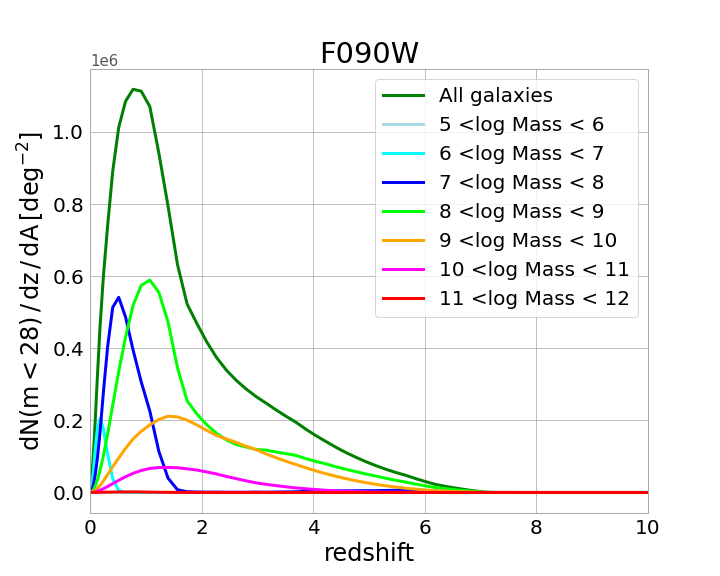}{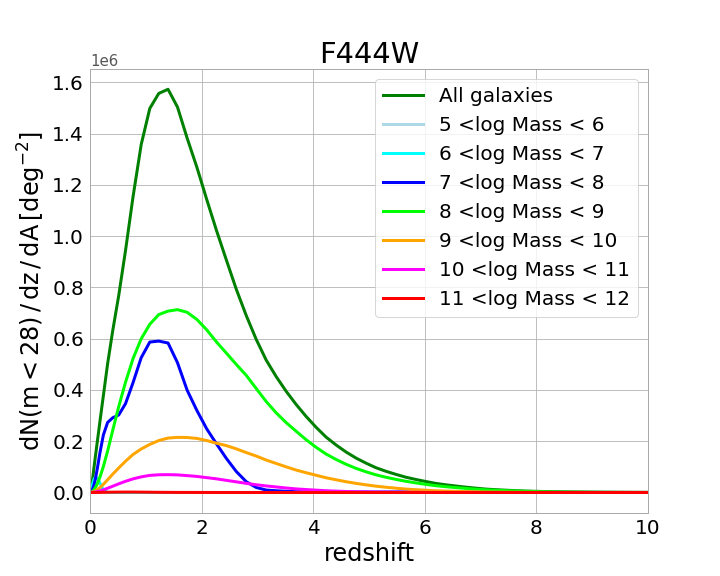}
\caption{Galaxy number counts (top row) and relative redshift distributions (bottom row) split in bins of stellar mass to study the contribution of each stellar mass bin to each apparent magnitude. The analysis is repeated for the shortest wavelength NIRCam filter available in PEARLS, F090W (left column) and the longest wavelength F444W. The dark green line in all panels includes all the simulated objects while other colours show the contribution of different stellar mass bins.
%it's the bright slope that is actually changing while the faint is somewhat the same. 
\label{fig:MASS}}
\end{figure*}

\subsection{Contribution of different stellar masses}
A further piece of information that can help in understanding the shape of the number counts comes from the stellar masses of galaxies. 
The question we want to address in Fig.~\ref{fig:MASS}, is which stellar masses are contributing at different apparent magnitudes to the number counts. A further question is if this contribution is significantly different in different NIRCam filters, i.e. if there is a wavelength dependence. 
In the top row of Fig.~\ref{fig:MASS} we show the number counts for the shorter (F090W) and longer (F444W) NIRCam filters available in the PEARLS observations used here. 
The overall counts (green line) are then split in different mass bins 1 dex wide and ranging from $10^5$ to $10^{12}\, M_\sun$.
This figure is somewhat related to Fig.~\ref{fig:counts_absmag} as absolute magnitudes are to an extent (depending on the wavelength)related to stellar masses (the more massive the galaxy is the more luminous it tends to be). 
In fact, we can see that the most massive galaxies, those with stellar masses between $10^{11} < M_*/M_{\odot} < 10^{12}$, are not those that contribute the most to the bright end of the counts but instead galaxies between $10^{10} < M_*/M_{\odot} < 10^{11}$ are dominant. 
This is in agreement with what we see in Fig.~\ref{fig:counts_absmag} where the brightest galaxies ($M_{\rm{AB}}<-23$) are not able alone to fully populate the bright end as lower luminosity galaxies (approximately $M_{\rm{AB}}<-21$) needs to be taken into account to fully reproduce the bright end of the counts. 
By comparing the two panels in the top row of Fig.~\ref{fig:MASS}, it is interesting how the contributions of different stellar masses are different when comparing  the shorter wavelength filter at $0.90 \,\mu$m (left panel) with the longer wavelength filter at $4.44 \,\mu$m (right panel).   
In the F090W filter, we can see that the contribution of different stellar masses spreads over a wide range of apparent magnitudes with a smooth profile that is reflected in a weak break in the slope of the overall counts. The knee in the counts for the filter F444W is definitely more pronounced as the contributions to the counts from the stellar mass selected samples are more peaked. Moreover, it is clear how the bright part of the counts, brighter than the break, is mainly constituted by galaxies with stellar masses bigger than $10^{10} M_{\odot}$. 
The last thing to note is that this simulation has been built with a dark matter halo mass resolution of $10^{8.5} M_{\odot}$ which for this model translates to a stellar mass resolution of about $10^{5} M_{\odot}$. 
However, from the top row of Fig.~\ref{fig:MASS}, we can see that galaxies with stellar masses below $10^6 M_{\odot}$ are not important in shaping the number counts as they would a very tiny number with an observational magnitude limit of $m=28$.
In the lower panels of Fig.~\ref{fig:MASS}, we see the corresponding redshift distributions, using the same colour code for the stellar mass bins. 
It can be seen that although the counts in the shorter wavelength filter have a broader distribution in apparent magnitude, the corresponding redshift distributions are more peaked than is the case with the longer wavelength filter. 
This shows how the relative shift of the single luminosity functions between themself is more important than the overall shift of the luminosity functions due to the cosmological dimming when interpreting the shape of the number counts. We devote the following section to the explanation of this statement.

\subsection{Relation between SED and number counts}
\label{sec:SED}

To understand the origin of the break in the number counts and why the change in slope is more pronounced at different wavelengths, it is important to understand how the region of the spectral energy distribution (SED) sampled changes as photons travel from the rest frame of the galaxy to the telescope (the observer frame). 
It is clear from Fig.~\ref{fig:LF} that the conversion between rest frame (left panel) to the observer frame (right panel) is the cause of the spread in the LFs with redshift and hence the break in the counts slope.
The reason why the break depends on the observed wavelength of the counts is related to the shape of the SED and its transition from the rest to the observer frame. 
The $k$-correction quantifies the transformation between the rest and observer frame. 
The $k$-correction takes into account that the observed flux $f_{\nu_o}$ is seen at $\nu_o$ (observer frame) while the emitted luminosity $L_\nu$ is emitted at a different frequency $\nu$ (rest frame): 
%In fact this can be clearly seen by comparing the two panels of Fig.~\ref{fig:LF}. By converting the absolute magnitudes in the rest frame of the galaxy (the units used in the luminosity functions) into apparent magnitudes in the observer frame (the units used in the number counts) we can see how the self-similar luminosity functions in the left panel spread over a wider range of magnitudes in the right panel. 
%This is where the broken knee is created as the procedure to obtain the number counts from a set of luminosity functions is to integrate them together over the window of redshift covered by the observations.
%We know that the relation between the observed flux in the observer frame, $f_{\nu_0}$, and the intrinsic luminosity of the galaxy in the rest frame $L_{\nu}$, is the following:
\begin{equation}
\label{eq:flux}
f_{\nu_0} = \frac{L_{\nu}}{4 \pi D_L^2}(1+z),
\end{equation}
%where $\nu_0$ is the redshifted frequency of the photons as measured in the observer frame and $\nu$ is the frequency of the emitted photons in the rest frame, 
where $\nu = (1+z) \nu_0$ \footnote{which implies $d\nu = (1+z) d\nu_0$ describing the bandwidth of the filters changing with redshift, and responsible for the $(1+z)$ term appearing in Eq.~\ref{eq:flux}.}.
Instead of using the luminosity in the rest frame, we can use the luminosity in the observer frame $L_{\nu_o}$ and incorporate all the other terms in the $k$-correction:
\begin{equation}
\label{eq:flux_Kz}
f_{\nu_0} = \frac{L_{\nu_0}}{4 \pi D_L^2} \cdot k(z,\nu,\nu_o), 
\end{equation}
with:
\begin{equation}
    k(z,\nu,\nu_o) = \left[ \frac{L_{\nu}}{L_{\nu_0}}(1+z) \right] .
\end{equation}
By taking the logarithm, we can convert everything into magnitudes where the observed flux converts into observer frame apparent magnitudes, $f_{{\nu}_o} \rightarrow m_{\rm{obs}}$, the rest frame luminosity converts into the rest frame absolute magnitude, $L_\nu \rightarrow M_{\rm{rest}}$, and the observer frame luminosity into the observer frame absolute magnitude, $L_{\nu_o} \rightarrow M_{\rm{obs}}$.
Eq.~\ref{eq:flux} becomes:
\begin{equation}
m_{\rm obs} = M_{\rm rest} + 25 + 5\log_{10}\left(\frac{D_L}{\rm Mpc}\right)-2.5 \log_{10}\left(1+z\right),
\end{equation}
while Eq.~\ref{eq:flux_Kz}:
\begin{equation}
\label{eq:mobs}
m_{\rm obs} = M_{\rm obs} + 25 + 5\log_{10}\left(\frac{D_L}{\rm Mpc}\right) + \mathcal{K},
\end{equation}
where:
\begin{equation}
    \mathcal{K} = -2.5 \log_{10}\left(1+z\right)-2.5 \log_{10}\left(\frac{L_\nu}{L_{\nu_o}}\right).
\end{equation}
It is convenient for our purpose to view the $k$-correction as being split into a redshift part and a frequency part:
\begin{equation}
    \mathcal{K}(z,\nu,\nu_o) = \mathcal{K}(z) + \mathcal{K}(\nu,\nu_o), 
\end{equation}
with the redshift component being:
\begin{equation}
  \mathcal{K}(z) =  -2.5 \log_{10}\left(1+z\right) ,
\end{equation}
and the frequency component being:
\begin{equation}
\label{eq:kcorr_nu}
    \mathcal{K}(\nu,\nu_o) = -2.5 \log_{10}\left(\frac{L_\nu}{L_{\nu_o}}\right) .
\end{equation}
This is very important as here we want to show that the feature responsible for the change in the intensity of the break in the counts is the frequency part of the $k$-correction, $\mathcal{K}(\nu,\nu_o)$ from Eq.~\ref{eq:kcorr_nu}, when applied to the infrared SED of a galaxy. 
%Notice that in the unrealistic case of a completely flat SED, there would be no frequency component for the $k$-correction as $L_\nu$ would be equal to $L_{\nu_o}$.
To study this, it is particularly convenient to study the difference between the observer and rest frame absolute magnitudes as it corresponds exactly to the frequency part of the $k$-correction:
\begin{equation}
\label{eq:Mrest_Mobs}
    M_{\rm obs} - M_{\rm rest} = -2.5\log_{10}\left(\frac{L_{\nu}}{L_{\nu_0}}\right) \equiv  \mathcal{K}(\nu,\nu_o).
\end{equation}
The advantage of using the GALFORM simulation is that we have access both to $M_{\rm rest}$ and $M_{\rm obs}$. This is something that is not possible by using only observational data.
In the unrealistic case in which the SED of a galaxy were flat (i.e. same intensity at all wavelengths), the intrinsic luminosity of a galaxy would be the same in the rest and observer frame, and there would be no difference between $M_{\rm rest}$ and $M_{\rm obs}$. 
However, for any SED with an intrinsic shape, $L_\nu$ would always be different from $L_{\nu_o}$ as we are probing different parts of the SED. 

\begin{figure*}%[h]
\plottwo{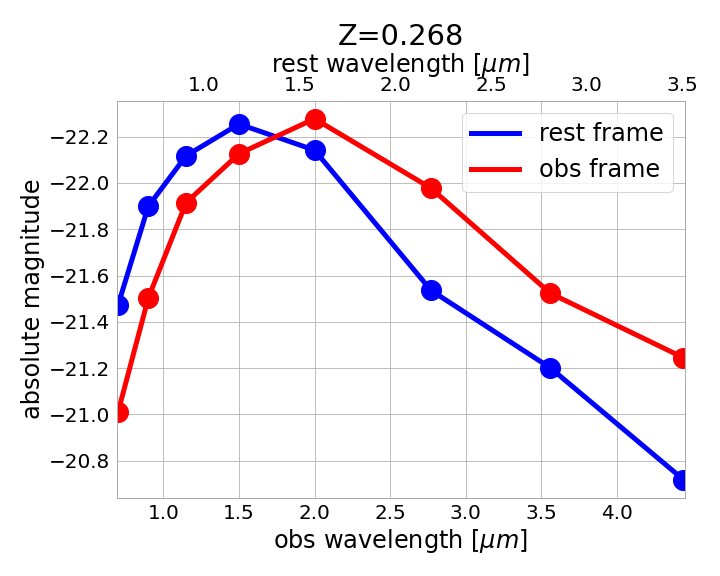}{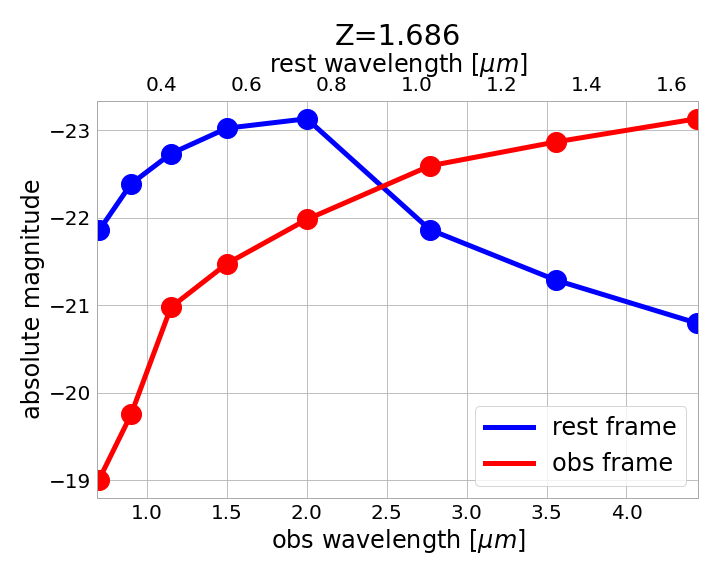}
\caption{Comparison between the observer (red line) and rest frame (blue line) SEDs of two example galaxies at different redshifts. The difference between the rest and observer frame is the frequency term of the $k$-correction as in Eq.~\ref{eq:Mrest_Mobs} (see text for more details). This figure shows how the band shifting of the SED can create a decrease in flux at the wavelengths before the crossover (positive $k$-correction) and an increase in flux for longer wavelengths (negative $k$-correction). At different redshifts the crossing point between the rest frame and the observer frame will be at different wavelengths affecting differently the behaviour of the different filters. The top axis shows wavelengths that the observed SED is actually sampling in the rest frame. The example galaxies were chosen to have a $B/T$ close to 1, in order to sample elliptical galaxies but notably, in these infrared bands, SED for galaxies with low $B/T$ seems to have the same shape of SED when plotted in absolute magnitudes instead of physical units like $F_{\lambda}$ or $F_{\nu}$.
\label{fig:SED}}
\end{figure*}
We now use Fig.~\ref{fig:SED} to understand how the frequency part of the $k$-correction can give rise to the break in the number counts. 
In this figure, we plot
%To understand this process we can use Fig.~\ref{fig:SED} where we plot 
$M_{\rm{obs}}$ with a red line and $M_{\rm{rest}}$ with a blue line, as a function of frequency, for two example galaxies at different redshifts. 
The rest frame SED would only overlap with the observed SED if the galaxy were at redshift zero.

We can see from Eq.~\ref{eq:Mrest_Mobs} how the difference between these two curves represents the frequency-dependent part of the k-correction. 
The key point is that in the transformation between rest and observer frame, this correction affects the flux differently in different filters, all the other terms in Eq.~\ref{eq:mobs}, used to get the apparent magnitudes for the number counts, depend only on redshift, so they have the same effect in all the filters.

The lower x-axis shows the wavelength of the NIRCam filters that range between $0.7\,\mu$m and $4.44\,\mu$m. 
This is the window of wavelengths that we can see in the observer frame with this instrument. 
On the top x-axis, we show the emitted wavelength in the rest frame.
If the galaxy were at redshift zero, the emitted and the observed frequency would be the same and as a consequence the emitted and observed SED would overlap. 
However, when we observe a galaxy at a higher redshift the SED that we observe (red line) appears redder than what it was originally emitted in the rest frame of the galaxy (blue line). 
This means that without a model SED we do not know the original shape of the SED at the rest wavelength indicated by the filter but we only know what the SED looks like at bluer rest wavelengths (as reported on the top axis).
%, i.e. the wavelength of the emitted photons in the rest frame of the galaxy before travelling to the observer frame).
%shifts to redder wavelengths (red line) while the filters keep looking at the same wavelengths so that they are actually probing different wavelengths as reported in the axis on the top of the figure (which show the wavelengths of the photons as they were emitted by the galaxy).
As a consequence, the process of converting magnitudes from the observer to the rest frame involves knowledge of the shape of the SED outside of the wavelengths of the filters used for observations. 
In the left panel of Fig.~\ref{fig:SED} it is easy to recognise that the conversion between the rest and the observer frame is just a shift that preserves the shape of the SED. 
However, as redshift increases the observer frame is probing a totally different window in wavelength and the shape of the rest frame SED is quite different from what is observed. 
That is why we cannot just shift the SED rigidly to other wavelengths but we need to correct for the difference between the red and the blue lines which is the ratio of the emitted luminosity and the observed luminosity, ${L_{\nu}}/{L_{\nu_0}}$ (see Eq.~\ref{eq:Mrest_Mobs}). 
In this sense, the frequency-dependent part of the $ k$-correction (Eq.~\ref{eq:Mrest_Mobs}) is used to change the shape of the SED to what it would look like if the galaxy were at redshift zero, so that we could actually see the galaxy in the rest wavelengths of the filters used.
This correction is crucial in the interpretation of the shape of the counts as depending on the redshift (i.e. the amount of shifting of the SED), the flux could be increased or decreased as described by the frequency part of the $k$-correction. 
In fact, looking at Fig.~\ref{fig:SED} we could see that on the left hand side of the crossing between the red and blue line, the conversion from rest frame to observer frame will create a decrease in flux while on the right-hand side it would create an enhancement (as the red line would have more flux than the blue for a fixed frequency). 
This effect is visible in both panels of Fig.~\ref{fig:SED}, although we can see that the crossing point between the two lines is in a different position and the extent of the $k$-correction is different. 
The $k$-correction is positive when it creates a decrease in the flux (by going from rest frame to observer frame) while the $k$-correction is negative when it creates an increase in the flux (right side of the crossing point). 

The key fact is that the effect of the $k$-correction is very different (it can create an increase or decrease in flux) at different wavelengths (i.e. different filters). 
This is what indirectly makes the break in the number counts more or less pronounced in different filters. 
To understand how this is happening, we can use Fig.~\ref{fig:LF}. 
If we ignore $\mathcal{K}(\nu,\nu_o)$, the transition between the luminosity functions in the left panel (rest frame) to the right panel (observer frame) would be the same in every filter as it would depend only on redshift. 
%rigid in the sense that the cosmological dimming would shift the luminosity functions to the same amount towards fainter magnitudes as we increase the redshift. 
However, when we include the smaller effect of $\mathcal{K}(\nu,\nu_o)$, this creates an increase in flux in the luminosity function at some redshift and a decrease in flux at other redshifts, depending on where the frequency of the filter would lay in Fig.~\ref{fig:SED} (left or right of the crossing point between rest and observer frame SED, i.e. positive or negative $k$-correction). 
The fact that for the same filter, at different redshifts the $k$-correction could be very different (see Fig.~\ref{fig:kcorr_z} in Appendix~\ref{app}), creates a ``compression'' of the observer frame luminosity functions at the position of the break in the counts.
When integrating the observer frame luminosity functions at all the redshifts to get the number counts this contributes to shaping the strength of the break. 

It is important to clarify that although the $k$-correction, plays a crucial role in shaping the break of the power law in the number counts, it is not the only reason for the break. To verify this we included in Appendix~A an analysis that shows how the galaxy number counts would look like if we do not consider the effect of the $k$-correction. 
We found that the break would still be in the counts as a result of the break in the constituent luminosity functions, however, the break would look the same in all the filters. When negative, the $k$-correction makes the breaks more pronounced, when positive makes the break shallower. This effect is clear in Figs.~\ref{fig:counts_all_nircam}~and~\ref{fig:obsrest} (check also Fig.~\ref{fig:kcorr_z} to see the $k$-correction as a function of redshift and see for which filters it is negative and for which filters is positive).

%To summarise we can state that the origin of the break in the number counts is the different effect (boost or decrease in flux) that the $k$-correction can have at different redshifts because of the intrinsic shape of the SED. 
%By changing the filter the same phenomenon happens but the strength is different, i.e. less compression at shorter wavelengths (less broken knee) and more compression at longer wavelengths (more broken knee).
A final remark is that although this effect has never been studied in detail to justify the shape of the infrared galaxy number counts, in submillimeter galaxies the boost in flux due to the negative k-correction is a well-known phenomenon and it has been used to observe galaxies that would be too faint to be observed otherwise.

% I can try to be clear by adding the explanation that if I fix a filter (vertical line in the SED figure) the boost in flux would be bigger at high redshift than at low redshift when moving from rest to observer frame that in the luminosity functions figure is left to right panel.

%Since all the information we have from observations is from the observer frame and we don't have access to the rest frame, what we could infer is $L_{\nu_0}$ and not $L_{\nu}$ as in Eq.~\ref{eq:flux}. 

\section{Conclusions}
We have taken advantage of the incredible depth of JWST observations from \cite{windhorst23} to develop a deeper understanding of the double power law which is clearly present in the NIRCam galaxy number counts, ranging from $0.7\,\mu$m to $4.4\,\mu$m. 
The change in slope is more pronounced at longer wavelengths than at shorter wavelengths. We made use of the GALFORM semi-analytical model to understand the origin of the break. The model is able to reproduce the JWST counts in all of the filters, reproducing the location of the change in slope and the strength of the break quite accurately (Fig.~\ref{fig:counts} with the associated Fig.~Set~1 available online; see also the GALFORM predictions for JWST in \citealt{cowley18} and \citealt{Lu2024}). 
This means that the physics responsible for the break in the counts is included in the model, suggesting that we are not going to uncover new galaxy formation physics purely through the study of the number counts. 
We split the analysis of the counts into luminosity functions and we found out from Fig.~\ref{fig:redshift_filters} that the redshifts of the luminosity functions dominating the counts are relatively low ($z<2$) when limiting our model to the observational magnitude cut of about 28 (as in the PEARLS observations in \citealt{windhorst23}).
This rules out the possibility of using the galaxy number count to infer strong constraints on the early universe as for this magnitude limit, the galaxies shaping the counts are those at low and intermediate redshifts. 
By studying the contributions of the luminosity functions in the outputs of the simulation at instantaneous snapshots we were able to uncover the origin of the break and why it changes strength with wavelength. 
The break in the galaxy number counts does not arise due to a particular physical process or anything related to structure formation (other than the physics defining the break in the luminosity function) but is an observational effect due to the fact that the number counts are defined in the observer frame and not in the rest frame like the luminosity function. 
When looking at the luminosity functions contributing to the counts in the rest frame, we note that they are all self-similar apart from a small contribution due to the intrinsic evolution of galaxies and the change in star formation (see left panel of Fig.~\ref{fig:LF}). 
However, when we look at the luminosity functions in the observer frame (right panel of Fig.~\ref{fig:LF}), for a given filter, the $k$-correction contributes in different ways at different redshifts creating a ``compression'' at the location of the break. 
The reason for that is the intrinsic shape of the SED and the resulting $k$-correction that for a fixed redshift could be either positive (decrease the flux) or negative (increase the flux), depending on the filter used. 
%and how different is the window of rest frame wavelengths probed at different redshift. The ratio between the observer frame SED and the rest frame SED, i.e. the $k$-correction, could either boost or decrease the flux of the luminosity function depending on the redshift of that specific luminosity function. 
To better understand this process we took two examples of GALFORM-simulated galaxies at different redshifts and compared the rest and the observer frame SED (Fig.~\ref{fig:SED}), as the difference between them defines the $k$-correction. 
For shorter wavelengths than the intersection between rest and observer frame SED, the flux in the observer frame will be lower than the rest frame flux. 
For longer wavelengths instead the flux will receive a boost when converting from rest to observer frame. 
The location of the intersection between observer and rest frame SED is different at every redshifts as the k-correction is different. 
This means that a luminosity function in a specific filter will move to brighter or fainter magnitudes depending on the boost or decrease in flux regulated by a positive or negative $k$-correction. 
%have an inversion in the contribution of the k-correction (decreased vs boosted flux) at the redshift when the crossing point passes through the filter used.

The effect of a negative k-correction is something that has been known for submillimeter galaxies for a long time \citep{blain93}. In fact, \citealt{blain93} shows that the extent of the k-correction at high redshift in the submillimeter window could be so large (and negative, i.e. boost in the observed flux) that can make visible galaxies that are too faint to be observed in the rest frame wavelengths. For the first time here, we use this effect to explain the shape of the near-infrared number counts and the clear knee as observed by James Webb's NIRCam. 
This work proves the importance of the change in reference frame in interpreting the galaxy number counts. It shows the crucial contribution of the $k$-correction in shaping the number counts and defining its strength. We show that the origin of the break is a combination of different observational effects related to the change in reference frame in an expanding Universe combined with the break present in the luminosity functions. The role of galaxy evolution in the shape of the constituent luminosity functions is minor compare to the effect of the change in reference frame. We can consider self-similar luminosity functions in the rest frame and still obtain the break in the observed number counts as shown in Fig.~\ref{fig:LF}.

With this work, we put together for the first time a full consistent and physically motivated framework to describe why the number counts look this way in the near-infrared deep observation of JWST. 
%This work can be used in future studies that attempt to distinguish on different dark matter models based on galaxy statistics. 
Regarding the study of the early Universe using galaxy statistics, we warn the reader that extra care must be taken when dealing with high redshift faint galaxies as these have no role in defining the shape of the galaxy counts, at least to the observational magnitude limit of $m_{\rm AM}\lesssim 28$. 
Now that we are fully aware of the observational biases that result from the change in the reference frame, we have set the ground for the use of galaxy formation models to test different dark matter models, by changing the relying dark matter power spectrum. \\

%%%%
\figsetstart
\figsetnum{1}
\figsettitle{Set of complementary figures to Fig.~\ref{fig:counts}: Number counts predicted by the model and observed in the available NIRCam filters. The data in the backgrounds and their relative fits are from \cite{windhorst23}.}
\figsetgrpstart
\figsetgrpnum{1.1}
\figsetgrptitle{F070W}
\figsetplot{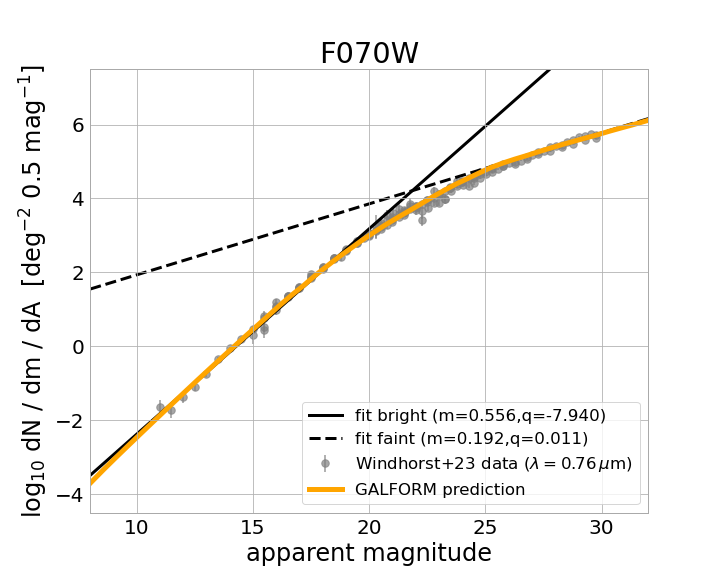}
\figsetgrpnote{Predicted counts for F070W, compared to data at $\lambda=0.76$.}
\figsetgrpend

\figsetgrpstart
\figsetgrpnum{1.2}
\figsetgrptitle{F090W}
\figsetplot{./figs/COUNTS_F090W_DATA.png}
\figsetgrpnote{Predicted counts for F090W, compared to data at $\lambda=0.88$.}
\figsetgrpend

\figsetgrpstart
\figsetgrpnum{1.3}
\figsetgrptitle{F115W}
\figsetplot{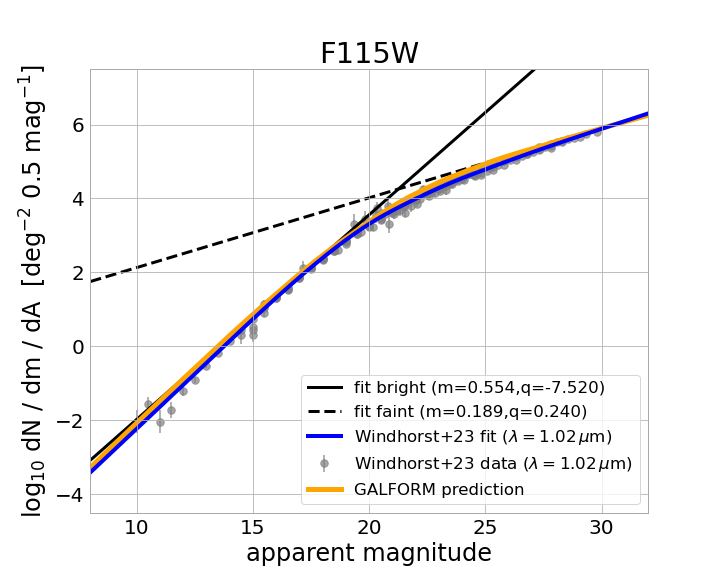}
\figsetgrpnote{Predicted counts for F115W, compared to data at $\lambda=1.02$.}
\figsetgrpend

\figsetgrpstart
\figsetgrpnum{1.4}
\figsetgrptitle{F115W}
\figsetplot{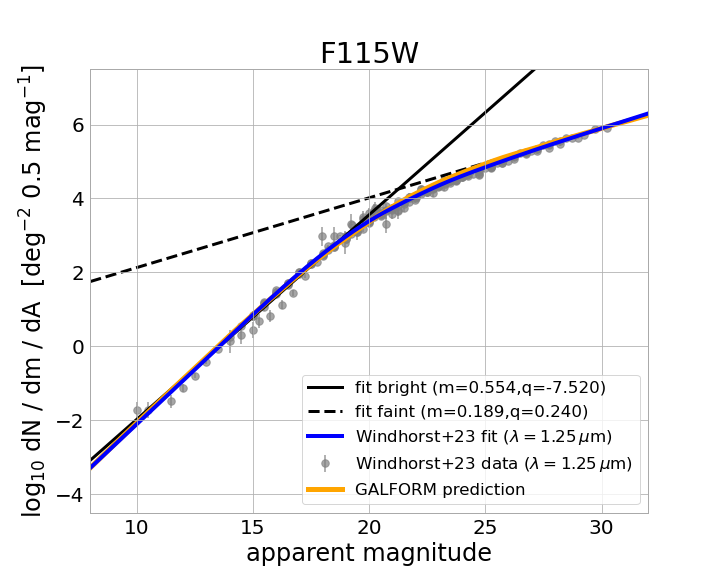}
\figsetgrpnote{Predicted counts for F115W, compared to data at $\lambda=1.25$.}
\figsetgrpend

\figsetgrpstart
\figsetgrpnum{1.5}
\figsetgrptitle{F150W}
\figsetplot{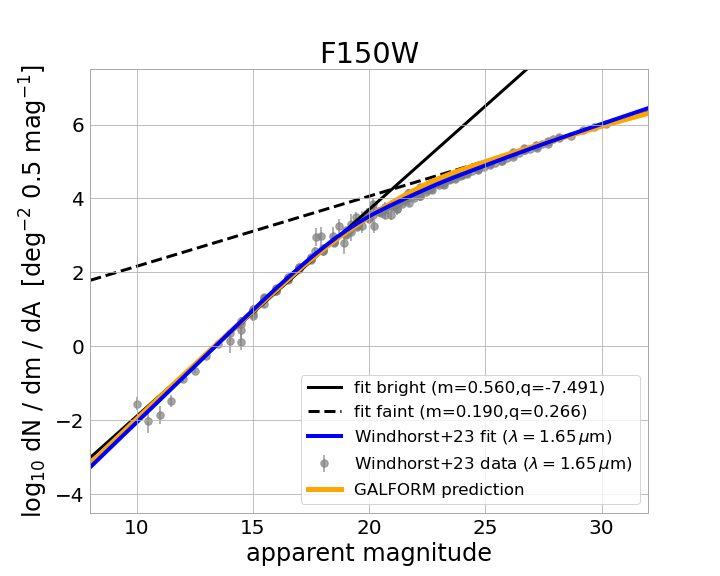}
\figsetgrpnote{Predicted counts for F150W, compared to data at $\lambda=1.65$.}
\figsetgrpend

\figsetgrpstart
\figsetgrpnum{1.6}
\figsetgrptitle{F200W}
\figsetplot{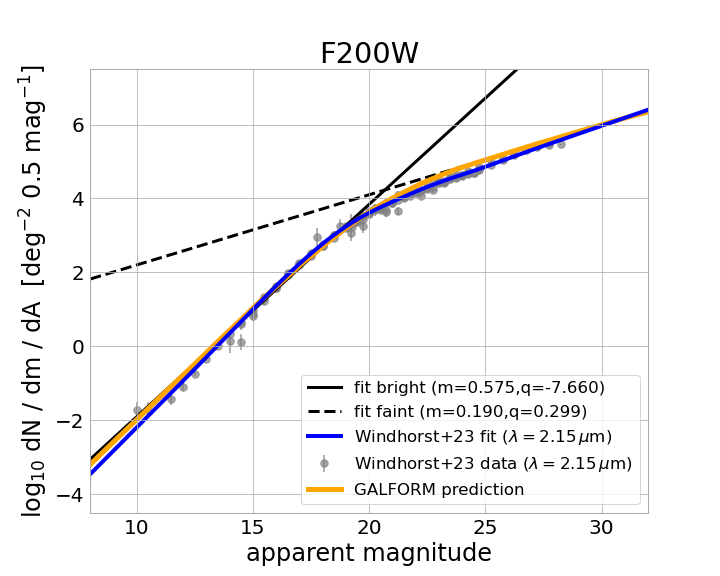}
\figsetgrpnote{Predicted counts for F200W, compared to data at $\lambda=2.15$.}
\figsetgrpend

\figsetgrpstart
\figsetgrpnum{1.7}
\figsetgrptitle{F356W}
\figsetplot{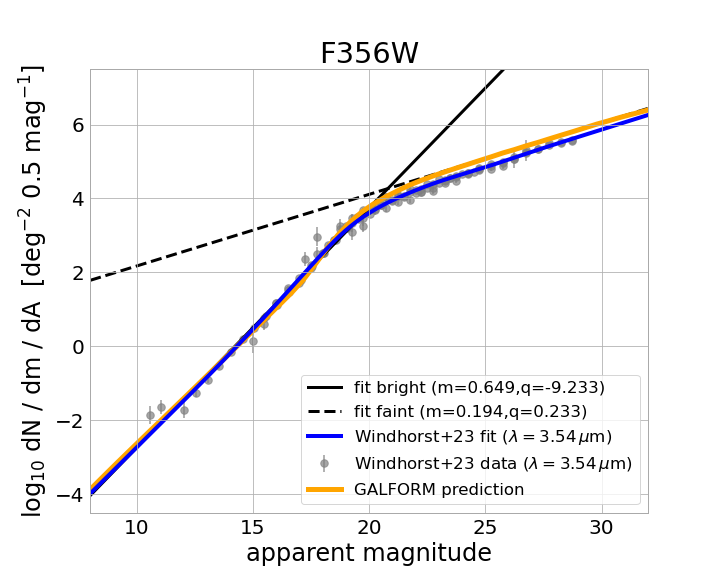}
\figsetgrpnote{Predicted counts for F356W, compared to data at $\lambda=3.54$.}
\figsetgrpend

\figsetgrpstart
\figsetgrpnum{1.8}
\figsetgrptitle{F444W}
\figsetplot{./figs/COUNTS_F444W_DATA.png}
\figsetgrpnote{Predicted counts for F444W, compared to data at $\lambda=4.49$.}
\figsetgrpend

\figsetend

%%%

\figsetstart
\figsetnum{2}
\figsettitle{Set of complementary figures to Fig.~\ref{fig:LF}: observer frame luminosity functions for all the NIRCam filters (analogue to the right panel of Fig.~\ref{fig:LF}). For longer wavelengths, the observer frame LFs span over a narrower range of apparent magnitudes resulting in a more pronounced broken knee. }
\figsetgrpstart
\figsetgrpnum{2.1}
\figsetgrptitle{F070W}
\figsetplot{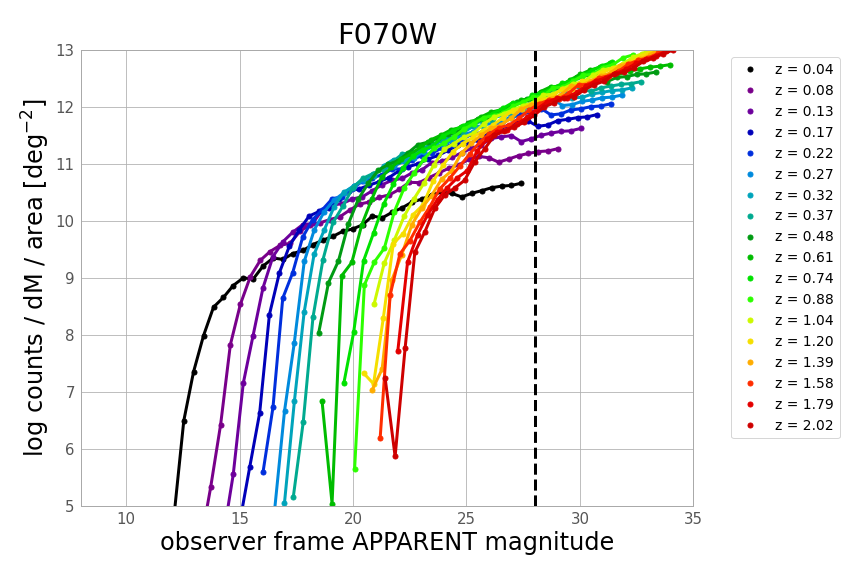}
\figsetgrpnote{Observer frame LFs for filter F070W.}
\figsetgrpend

\figsetgrpstart
\figsetgrpnum{2.2}
\figsetgrptitle{F090W}
\figsetplot{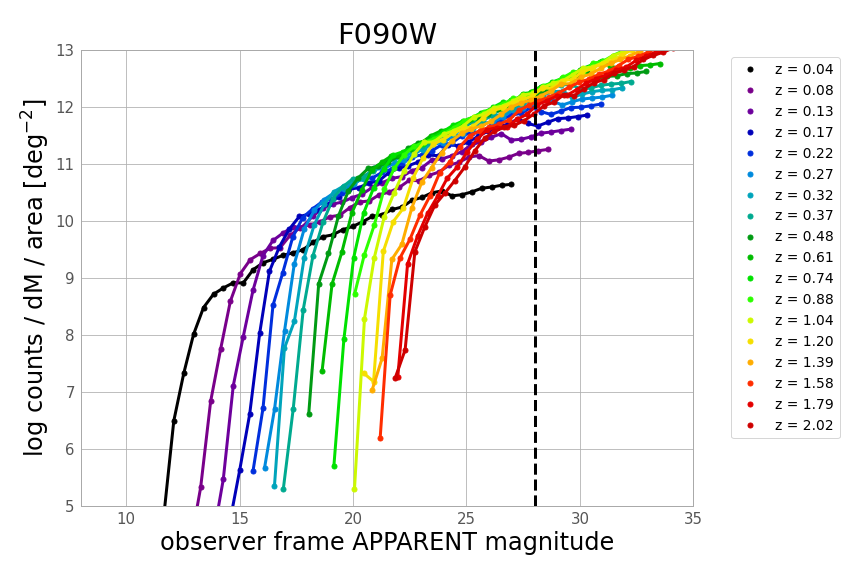}
\figsetgrpnote{Observer frame LFs for filter F090W.}
\figsetgrpend

\figsetgrpstart
\figsetgrpnum{2.3}
\figsetgrptitle{F115W}
\figsetplot{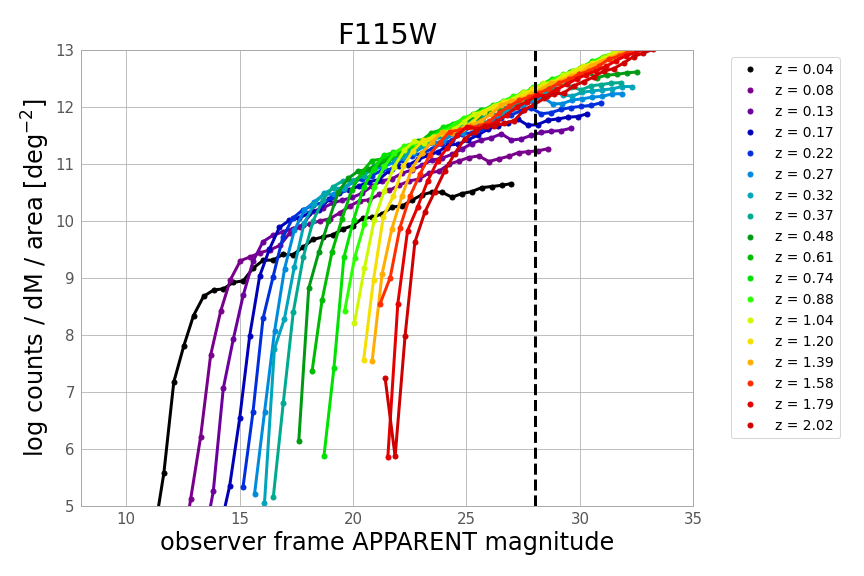}
\figsetgrpnote{Observer frame LFs for filter F115W.}
\figsetgrpend

\figsetgrpstart
\figsetgrpnum{2.4}
\figsetgrptitle{F150W}
\figsetplot{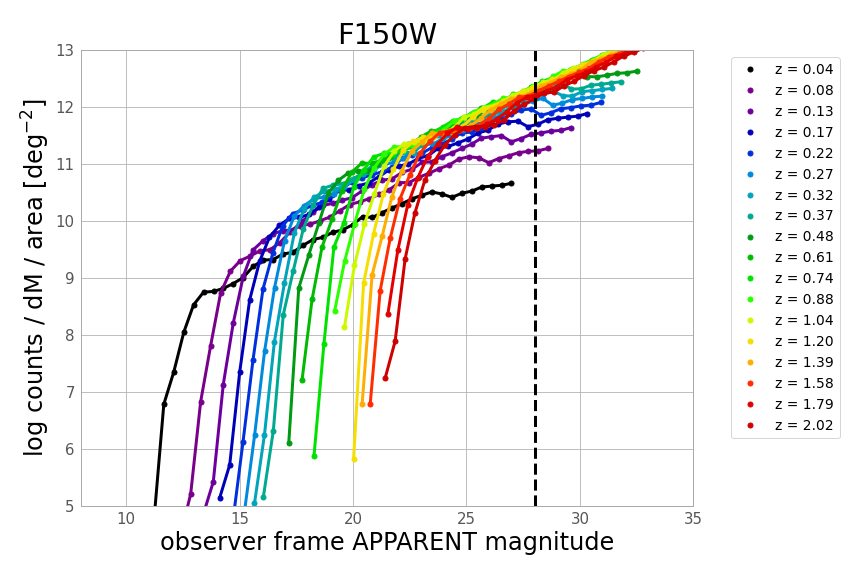}
\figsetgrpnote{Observer frame LFs for filter F150W.}
\figsetgrpend

\figsetgrpstart
\figsetgrpnum{2.5}
\figsetgrptitle{F200W}
\figsetplot{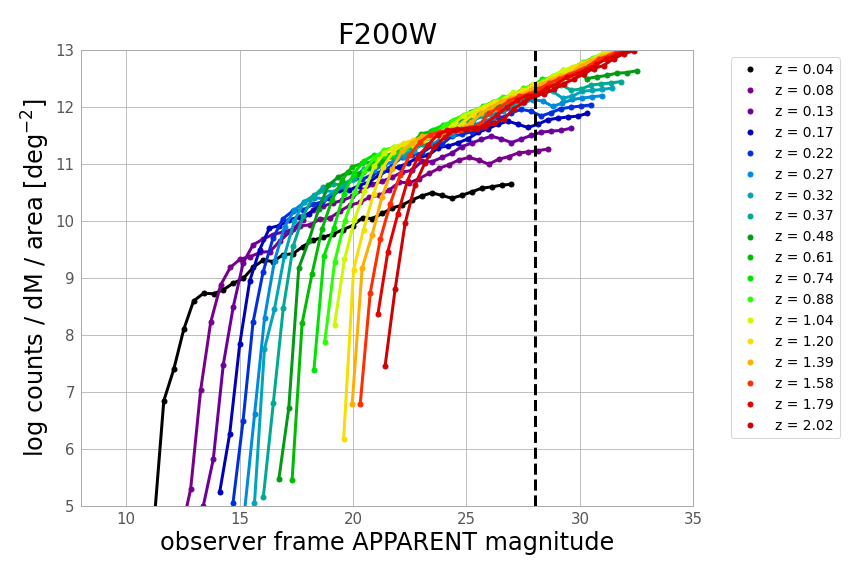}
\figsetgrpnote{Observer frame LFs for filter F200W.}
\figsetgrpend

\figsetgrpstart
\figsetgrpnum{2.6}
\figsetgrptitle{F277W}
\figsetplot{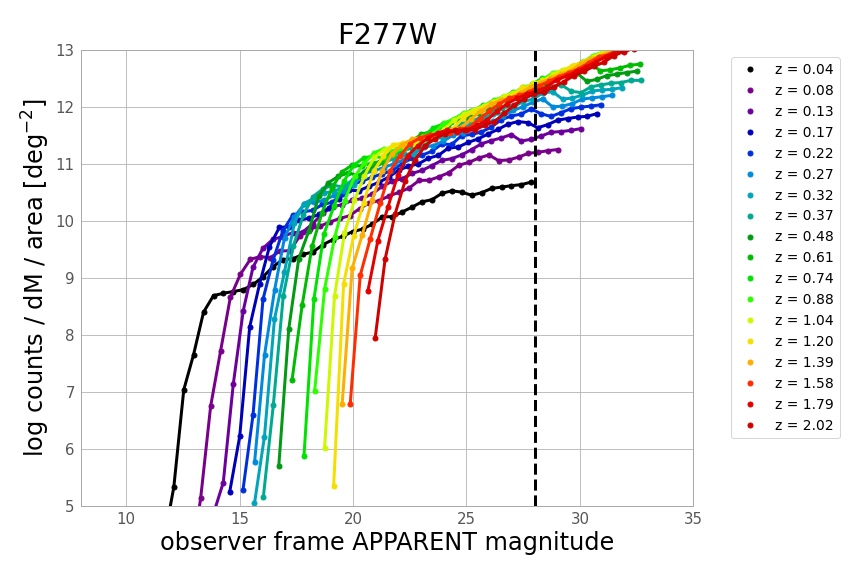}
\figsetgrpnote{Observer frame LFs for filter F277W.}
\figsetgrpend

\figsetgrpstart
\figsetgrpnum{2.7}
\figsetgrptitle{F356W}
\figsetplot{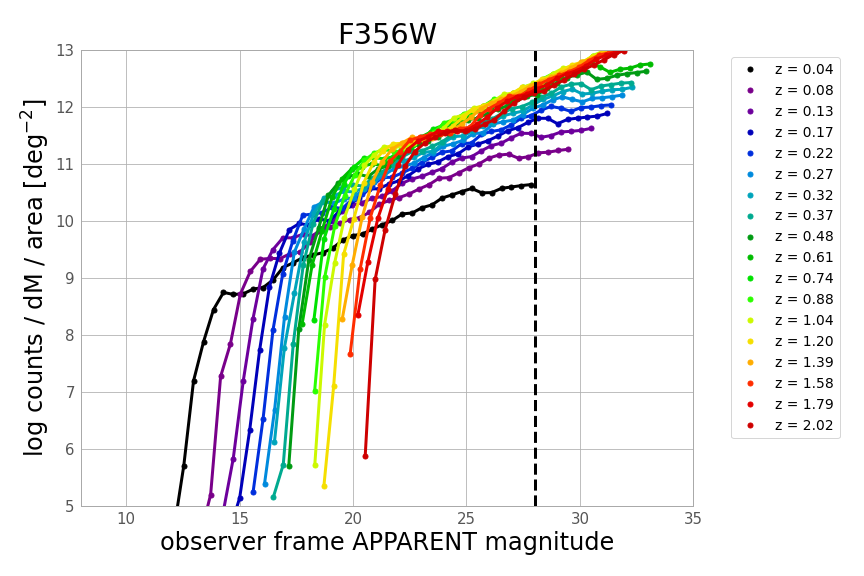}
\figsetgrpnote{Observer frame LFs for filter F356W.}
\figsetgrpend
 
\figsetgrpstart
\figsetgrpnum{2.8}
\figsetgrptitle{F444W}
\figsetplot{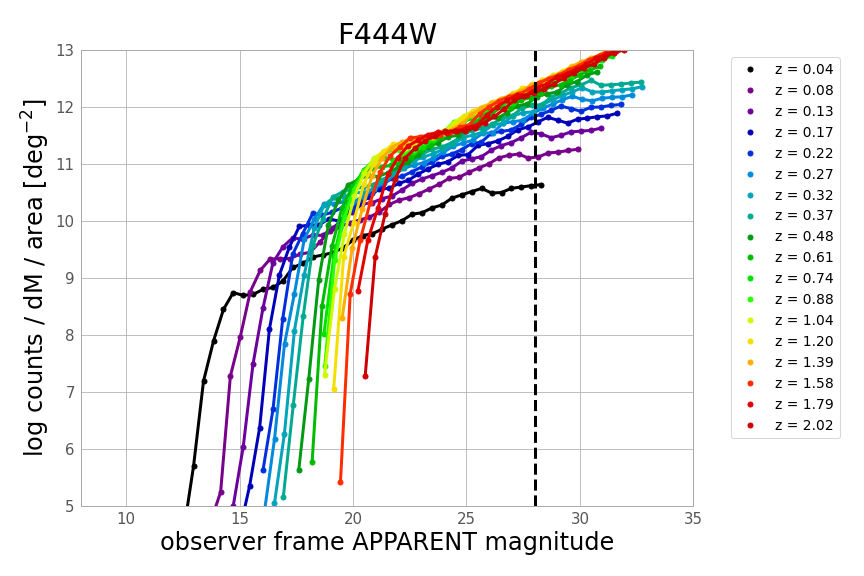}
\figsetgrpnote{Observer frame LFs for filter F444W.}
\figsetgrpend
 
\figsetend

\figsetstart
\figsetnum{3}
\figsettitle{Set of complementary figures to Fig.~\ref{fig:LFs_apprest}: rest frame luminosity functions in apparent magnitudes for all the NIRCam filters. Without the effect of the k-correction to account for the change in the reference frame, all the luminosity functions shift in the same way in all of the filter without the ``compression'' responsible for the stronger break at longer wavelengths seen in the observer frame.}
\figsetgrpstart
\figsetgrpnum{3.1}
\figsetgrptitle{F070W}
\figsetplot{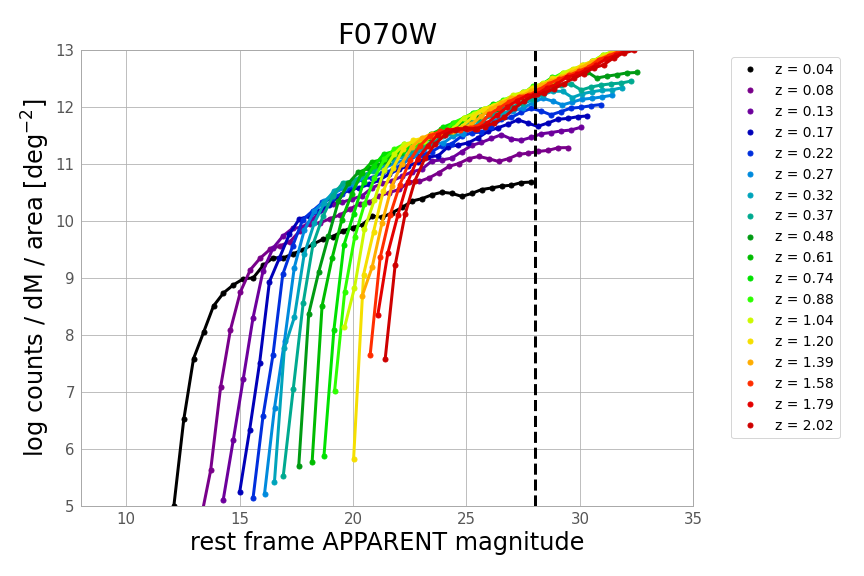}
\figsetgrpnote{Rest frame apparent magnitudes LFs for filter F070W.}
\figsetgrpend

\figsetgrpstart
\figsetgrpnum{3.2}
\figsetgrptitle{F090W}
\figsetplot{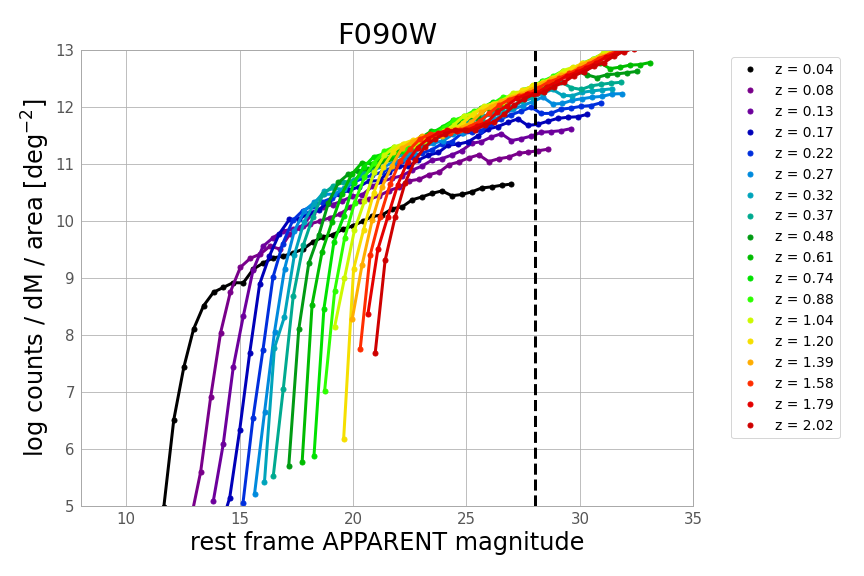}
\figsetgrpnote{Rest frame apparent magnitudes LFs for filter F090W.}
\figsetgrpend

\figsetgrpstart
\figsetgrpnum{3.3}
\figsetgrptitle{F115W}
\figsetplot{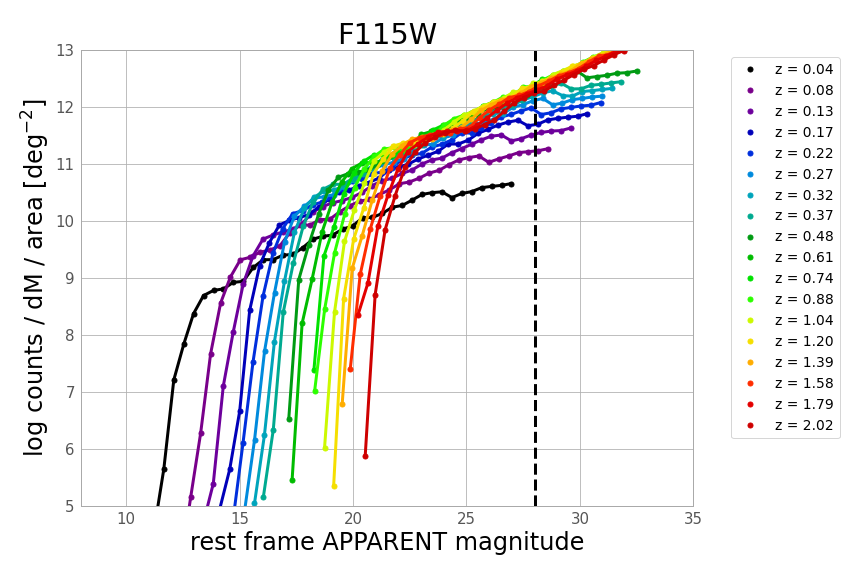}
\figsetgrpnote{Rest frame apparent magnitudes LFs for filter F115W.}
\figsetgrpend

\figsetgrpstart
\figsetgrpnum{3.4}
\figsetgrptitle{F150W}
\figsetplot{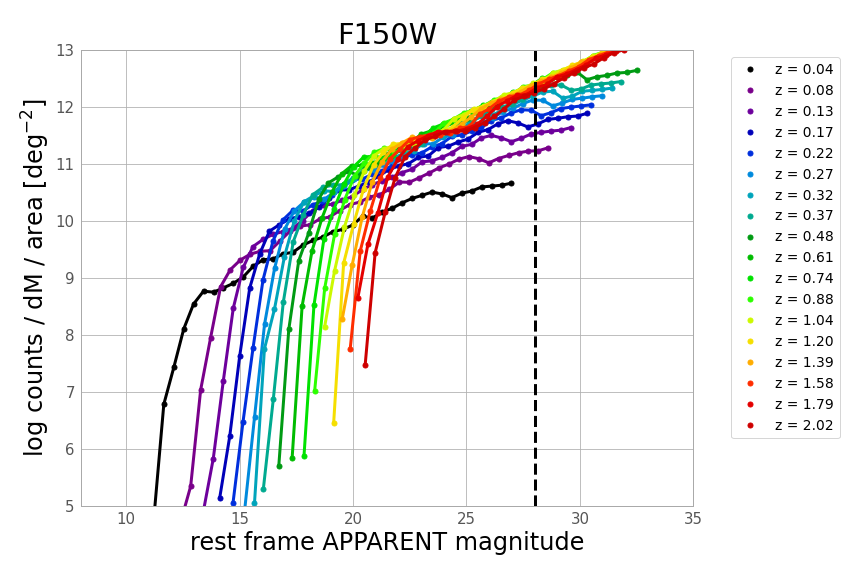}
\figsetgrpnote{Rest frame apparent magnitudes LFs for filter F150W.}
\figsetgrpend

\figsetgrpstart
\figsetgrpnum{3.5}
\figsetgrptitle{F200W}
\figsetplot{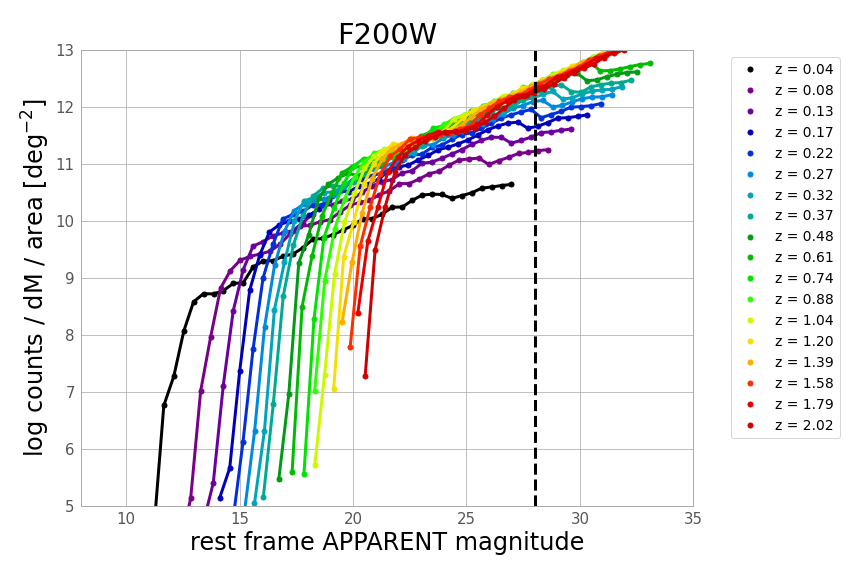}
\figsetgrpnote{Rest frame apparent magnitudes LFs for filter F200W.}
\figsetgrpend

\figsetgrpstart
\figsetgrpnum{3.6}
\figsetgrptitle{F277W}
\figsetplot{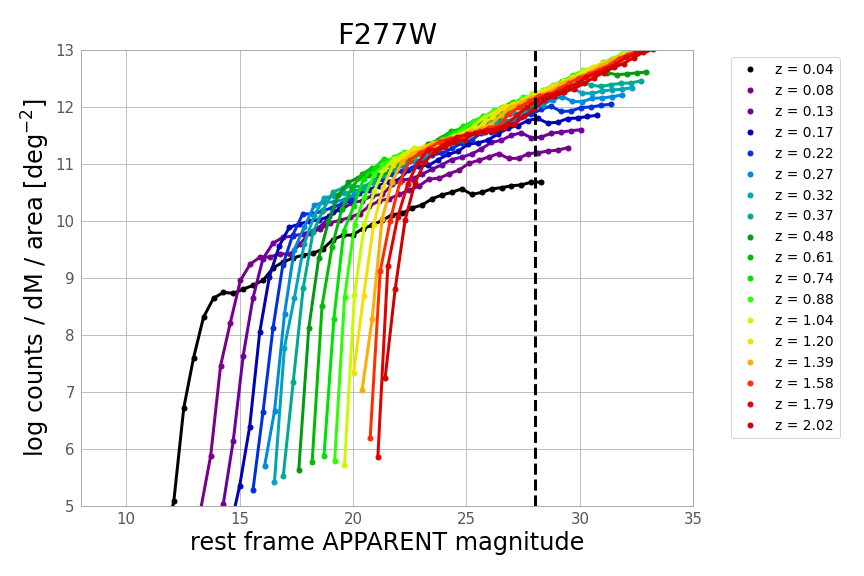}
\figsetgrpnote{Rest frame apparent magnitudes LFs for filter F277W.}
\figsetgrpend

\figsetgrpstart
\figsetgrpnum{3.7}
\figsetgrptitle{F356W}
\figsetplot{./figs/frame007.png}
\figsetgrpnote{Rest frame apparent magnitudes LFs for filter F356W.}
\figsetgrpend
 
\figsetgrpstart
\figsetgrpnum{3.8}
\figsetgrptitle{F444W}
\figsetplot{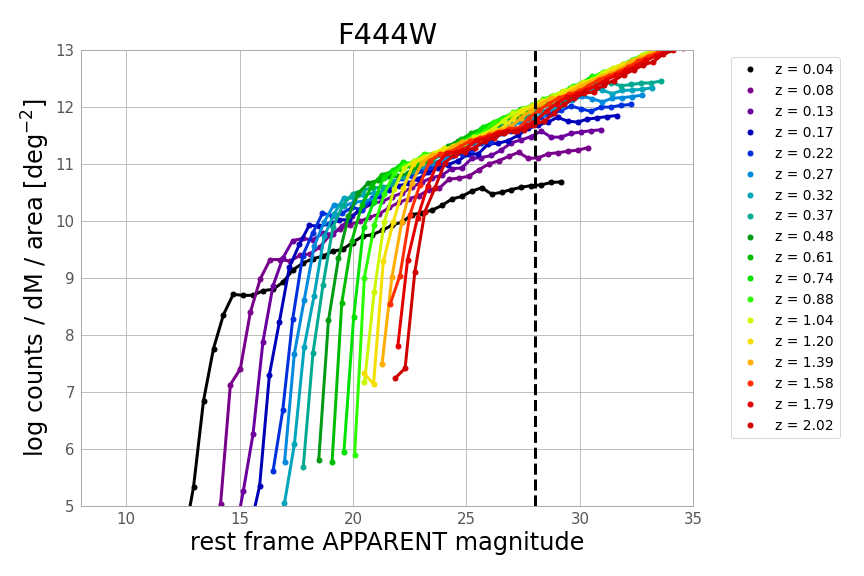}
\figsetgrpnote{Rest frame apparent magnitudes LFs for filter F444W.}
\figsetgrpend
 
\figsetend

\section*{Data Availability}
The data that support the findings of this study are stored at the Durham COSMA facilities. Please contact the corresponding author to have access.

\section*{acknowledgements}
GM is supported by the Collaborative Research Fund under Grant No. C6017-20G which is issued by the Research Grants Council of Hong Kong S.A.R.
This work used the DiRAC@Durham facility managed by the Institute for Computational Cosmology on behalf of the STFC DiRAC HPC Facility (www.dirac.ac.uk). The equipment was funded by BEIS capital funding via STFC capital grants ST/K00042X/1, ST/P002293/1, ST/R002371/1 and ST/S002502/1, Durham University and STFC operations grant ST/R000832/1. DiRAC is part of the National e-Infrastructure. 
This work is based on observations made with the NASA/ESA/CSA James Webb Space
Telescope. The data were obtained from the Mikulski Archive for Space
Telescopes at the Space Telescope Science Institute, which is operated by the
Association of Universities for Research in Astronomy, Inc., under NASA
contract NAS 5-03127 for JWST. These observations are associated with JWST
programs 1176 and 2738.

\appendix 
\section{The role of $k$-correction in shaping the counts at different wavelengths}
\label{app}

Given the conclusion of the paper on the importance of the $k$-correction in shaping the break in the power-law describing the number counts, we use this section to clarify that a break in the power-law would still exist in the number counts even without considering the $k$-correction but it would be a lot shallower at longer wavelengths as it is in the shorter wavelengths. 
Moreover, the intensity of the break would be exactly the same at all wavelengths. 
In the analysis carried out in this appendix, we study how the number counts would look like when we do not take into account the effect of the $k$-correction.

First we want to make a clarification on the terminology as it could easily bring to confusion.
We often refer to \textit{the effect of the $k$-correction} as the transition to different reference frames. 
The rest frame and the observer frame differ only by the amount called $k$-correction. 
Whether we add or subtract this amount is just the sign of the $k$-correction which is a convention. 
However, for observer astronomers that have access only to the observer frame, it is convention to say that we add the $k$-correction when converting from observer frame to rest frame and we remove the $k$-correction when we pass from rest frame to observer frame. 
Since the intrinsic properties of galaxies (i.e. what we ultimately want) are in the rest frame, you never need to remove the $k$-correction to go to the observer frame.
However in this work we do the opposite as we start from the prediction of the simulation that are the intrinsic properties in the rest frame (in our case the luminosity functions) and we remove the $k$-correction to see how observed properties look like in the observer frame (in our case the number counts). 
For this reason when we talk about \textit{the effect of the $k$-correction} we always refer to the removal of the $k$-correction or in other words, how properties look like when we don't know their intrinsic value but we are biased by the fact that we are forced to observe from the observer frame. 

Another source of confusion is when it comes to absolute and apparent magnitudes. 
The conversion between the two takes into account purely the distance of the object but there is not any correction for the fact that the frequency of the observed flux of the galaxy comes from a different part of the SED compared to when it was emitted, as this last part is taken into account by the $k$-correction. 

In general, it does make sense to study apparent magnitudes in the observer frame and absolute magnitudes in the rest frame as this would take into account both the ``de-distancing'' and the change in wavelengths due to redshift. 
However, in this analysis where we try to reconstruct step by step the observation starting from the simulated results, it is particularly beneficial to consider separately all the effects: conversion from absolute magnitude to apparent magnitude and conversion between rest frame to observer frame. \\

\begin{figure*}%[h]
\plottwo{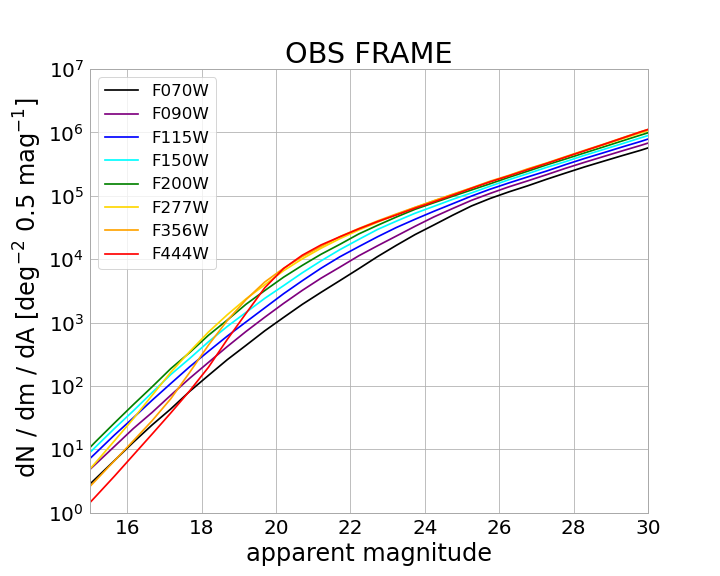}{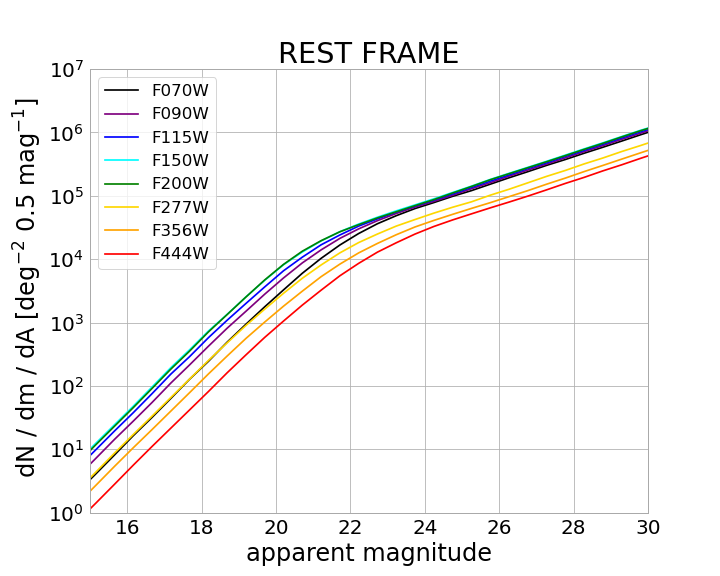}
\caption{Galaxy number counts in the observer (left panel) and rest frame (right panel) for all the Wide NIRCam filters.
\label{fig:counts_all_nircam}}
\end{figure*}

In this context, in Fig.~\ref{fig:counts_all_nircam}, we compare the number counts in the observer frame apparent magnitudes (the standard way in observational astronomy, in the left panel) and the number counts in the rest frame apparent magnitude (right panel). 
We can see that in the rest frame, the break in the power law is still there but the strength of the break is approximately the same, i.e. all the counts are approximately parallel to each other. 
The change to the observer frame, obtained by removing the $k$-correction from the rest frame, is what creates the difference in slope between the counts at different wavelengths and it is visible as well how the F444W get a steeper slope in the bright part compared to all the other counts, after receiving a boost in flux given by the negative $k$-correction. We have a negative $k$-correction when we receive a boost in flux by going from rest to observer frame. This means that intrinsically (in the rest frame) we would see fewer galaxies in the F444W filter, however because of the change to the observer frame, the flux gets enhanced and we see more. The opposite happens in the shorter wavelength filters like the F070W. Intrinsically, we would see more objects in the rest frame of F070W but moving to the observer frame, the positive $k$-correction decreases the flux and hence the number of objects observed.  
\begin{figure}
    \gridline{\fig{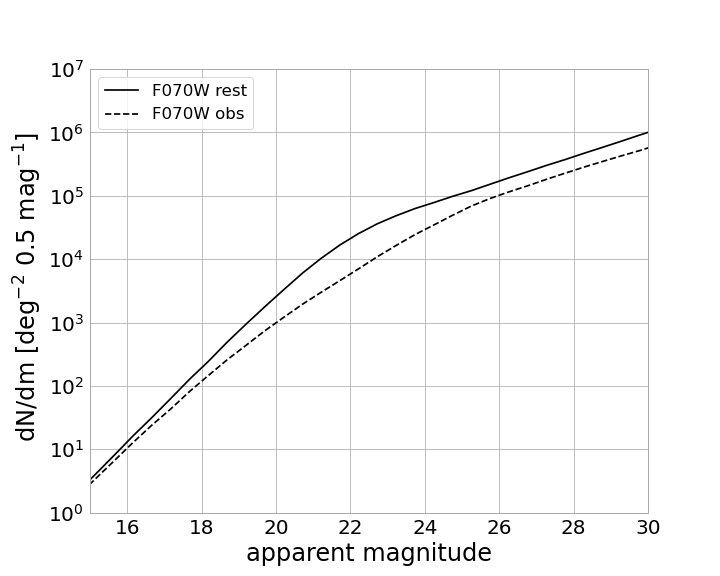}{0.40\textwidth}{\hspace{-20mm}}
                \fig{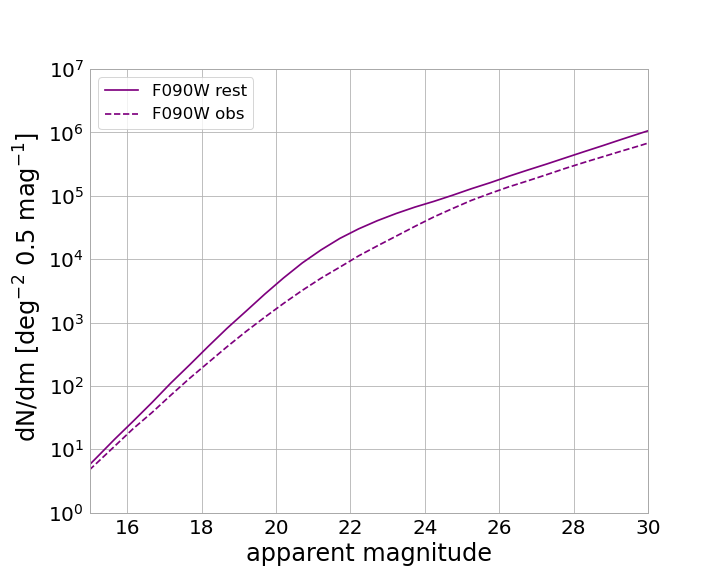}{0.40\textwidth}{}}
    \vspace{-10mm}
    \gridline{\fig{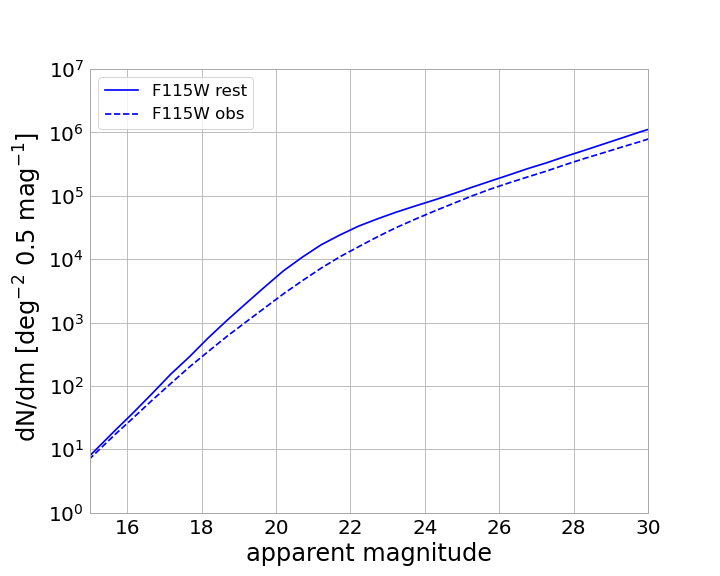}{0.40\textwidth}{\hspace{-20mm}}
                \fig{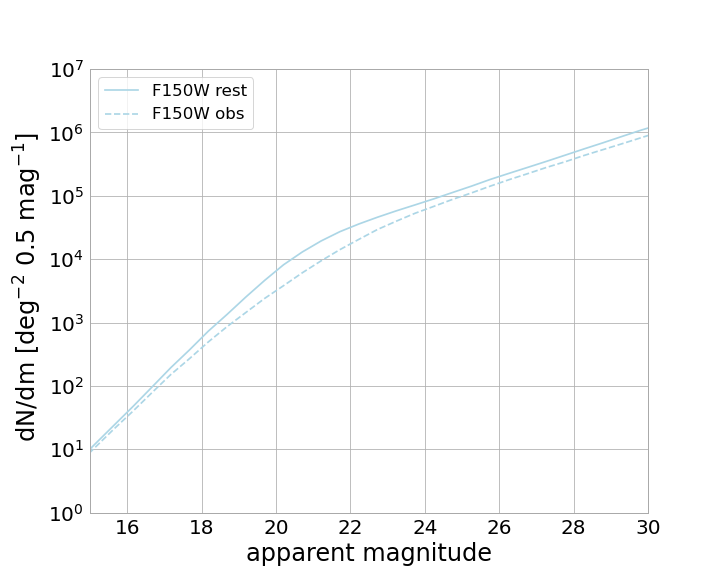}{0.40\textwidth}{}}
    \vspace{-10mm}
    \gridline{\fig{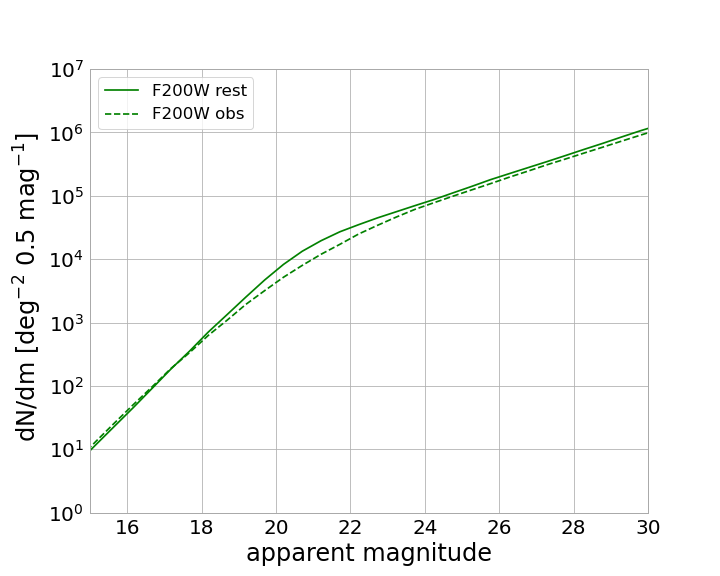}{0.40\textwidth}{\hspace{-20mm}}
                \fig{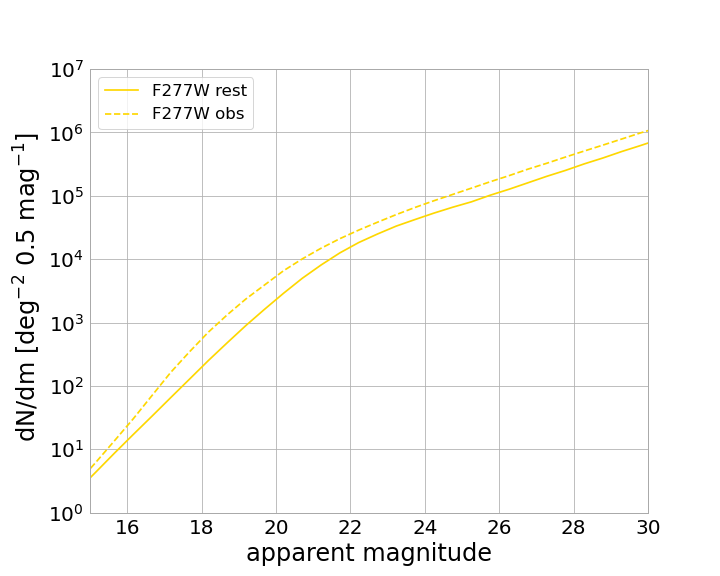}{0.40\textwidth}{}}
    \vspace{-10mm}
    \gridline{\fig{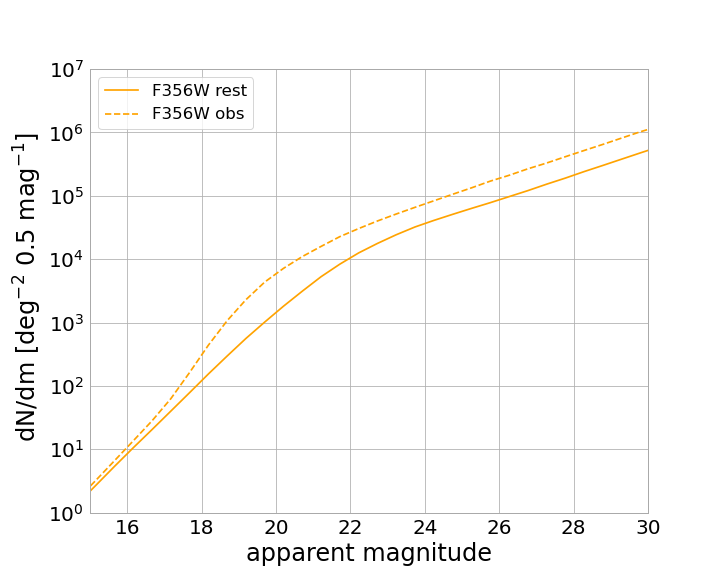}{0.40\textwidth}{\hspace{-20mm}}
                \fig{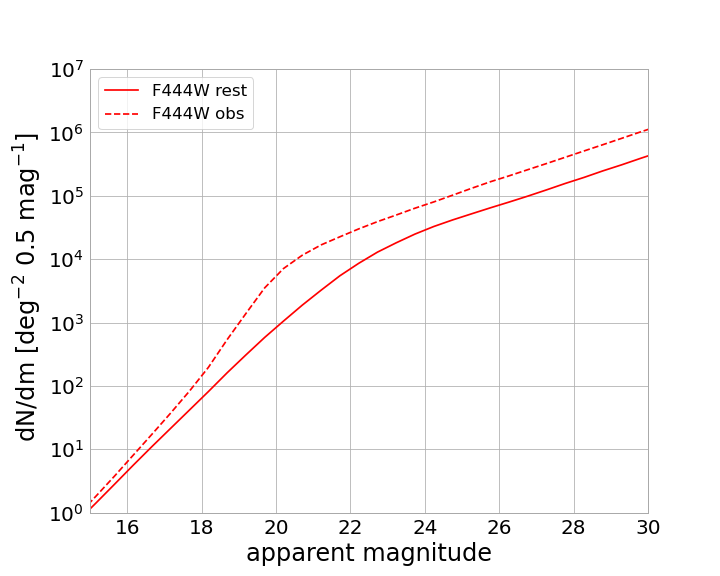}{0.40\textwidth}{}}
    \vspace{-4mm}
    \caption{Comparison between the number counts in the observer frame and in the rest frame to study the effect of the change in reference frame controlled by the k-correction.}
    \label{fig:obsrest}
\end{figure}

This is easy to see in Fig.~\ref{fig:obsrest} where we split Fig.~\ref{fig:counts_all_nircam} into different panels for each filter and we draw with a dashed line the observer frame apparent magnitudes (the standard) and with a solid line the rest frame apparent magnitude to see how the counts would look like without the change in reference frame obtained by the removal of the $k$-correction. 
Starting from F070W on the top left, we can see how the observer frame counts are fewer than the rest frame (positive $k$-correction). 
Going to increasingly longer wavelengths the observer frame counts approach the rest frame counts until this trend is reversed between F200W and F277W. 
This corresponds to the transition from positive to negative $k$-correction for most of the galaxy redshifts involved in the number counts. 

\begin{figure}
\centering
\scalebox{0.6}{\plotone{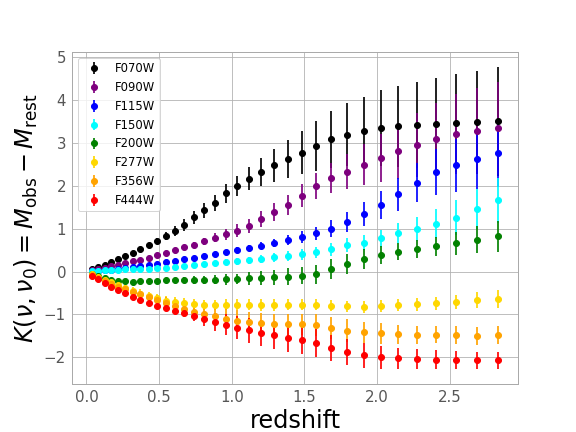}}
\caption{Frequency part of the k-correction as a function of redshift for all the Wide NIRCam filters, see Eq.~\ref{eq:kcorr_nu}. Median and standard deviation from the simulated sample at different snapshots.
\label{fig:kcorr_z}}
\end{figure}
To see this clearly, we plotted in Fig.~\ref{fig:kcorr_z} the $k$-correction as a function of redshift for all of the Wide NIRCam filters. 
We can see that shorter wavelengths than F200W have all positive $k$-correction (decrease in flux in the observer frame), while longer wavelength filters have all negative $k$-correction (increase in flux in the observer frame). 
The important information we get from Fig.~\ref{fig:kcorr_z} is that the $k$-correction is not the same at all redshifts but it becomes more important (in absolute value) at higher redshifts (see also Fig.~\ref{fig:SED} to understand why it is the case). The variation of the $k$-correction as a function of redshift is the reason of the stronger break at longer wavelengths and a shallower break at shorter wavelengths. 
To understand this, we can focus for example on the bottom right panel of Fig.~\ref{fig:obsrest} (filter F444W), we see that the bright part of the number counts in F444W are similar in the rest and observer frame. 
This is because the bright objects are dominated by low redshift galaxies with small $k$-corrections.
The faint part instead differs a lot between rest and observer frame and this is because faint objects are at higher redshifts and the effect of the $k$-correction is stronger. 
The transition between a negligible $k$-correction at low redshifts to a strong $k$-correction at high redshift is what shape the break in the number counts. 
For longer wavelengths where the $k$-correction is negative (boost in flux in the observer frame), it makes the break more pronounced while for shorter wavelengths were the $k$-correction is positive (decrease in flux in the observer frame), it smoothes the counts to a less pronounced break. \\

\begin{figure}
\centering
\scalebox{0.6}{\plotone{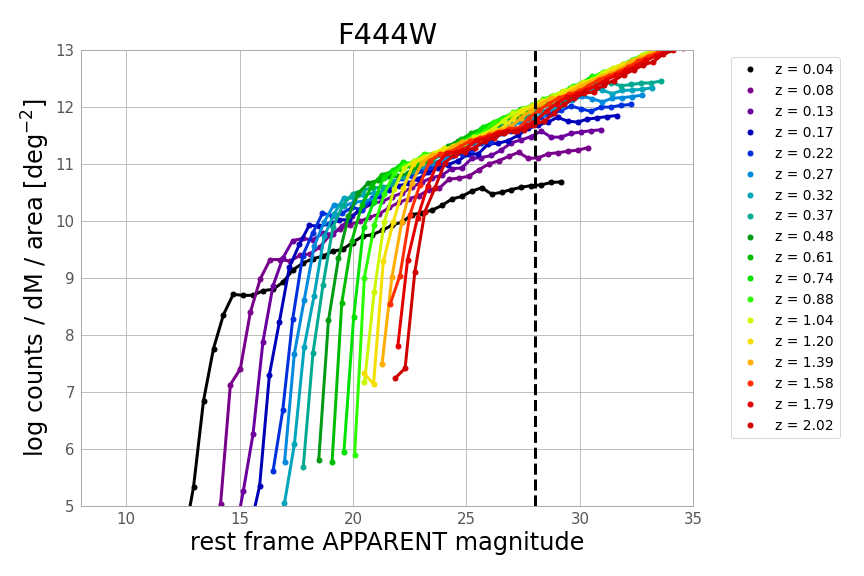}}
\caption{Constituent luminosity function in the rest frame apparent magnitude in filter F444W. We can see that without taking into account the change in the reference frame (the effect of the k-correction) the LFs do not compress around the break with a resulting break in the number counts that is smoother, see the solid line in the bottom right panel of Fig.~\ref{fig:obsrest}. See the other filters in Figure~Set~3, in the online version, and compare it with Figure~Set~2, which is the analogue in the observer frame (i.e. taking into account the effect of k-correction to change the reference frame). 
\label{fig:LFs_apprest}}
\end{figure}

Moving to the study of the constituent luminosity functions at specific redshfit outputs of the simulation, in the main text we have used Fig.~\ref{fig:LF}, and the associated Fig.~Set~2, to show how the change in the reference frame affects the LFs at different redshifts and hence the resulting galaxy counts. 
In Fig.~\ref{fig:LFs_apprest}, and the associated Fig.~Set~3, we want to see how the constituent luminosity functions would look like if we don't take into account the effect of the $k$-correction to correct for the change in reference frame. 
The LFs do not compress anymore around the break in the longer wavelengths, resulting in a shallower break in the number count. Comparing Fig.~Set~2 and Fig.~Set~3, in the online version it can be seen that while in the observer frame the LFs compress around the break at longer wavelengths, this does not happen when using the rest frame (i.e. without removing the $k$-corrections). In the rest frame the LFs at all redshift shifts together at different wavelengths preserving their relative apparent magnitude separation (and hence preserving the strength of the break at all wavelengths). 
This further proves that the change in the reference frame is responsible for a stronger break in the number counts at long wavelengths. \\

\begin{figure}
\centering
\scalebox{0.6}{\plotone{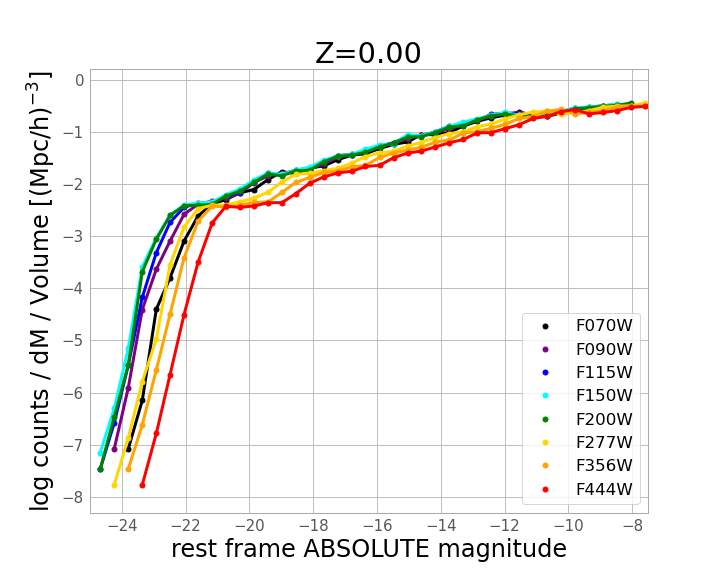}}
\caption{Luminosity functions in rest frame absolute magnitudes at $z=0$ for all the NIRCam Wide filters. The intrinsic luminosities of galaxies vary depending on the filter with a peak between F150W and F200W. This corresponds to the peak of the Integrated Galaxy Light (IGL) and, as a consequence, to the brightest break in the number counts.  
\label{fig:LFs_z0}}
\end{figure}

\begin{figure}
\centering
\scalebox{0.6}{\plotone{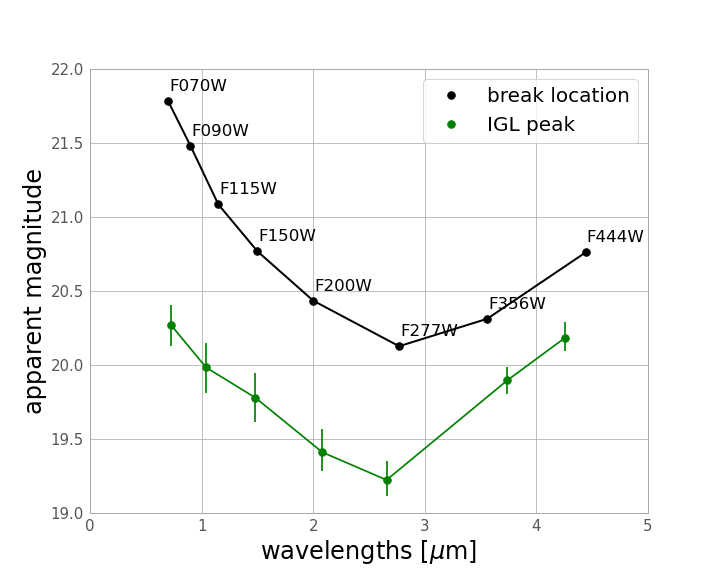}}
\caption{Apparent magnitude of the break in the number counts as a function of wavelengths (black line). The trend is a consequence of the intrinsic luminosity of the luminosity functions in Fig.~\ref{fig:LFs_z0}. The same trend is seen in the Integrated Galaxy Light (IGL) with a peak at around $2-3\,\mu$m.
\label{fig:break_IGL}}
\end{figure}

%%%%%%%%%%%%%%%%%%%%%%%%%%%%%%%%%%%%%
One question that arises naturally is the following. Given that the effect of the $k$-correction is responsible for the differences in the number counts in different filters, why do we still see differences in the right panel of Fig.~\ref{fig:counts_all_nircam} where we plot the number counts in the rest frame? 
Why do the overall number counts shift in the average apparent magnitude even in the rest frame? And why is this shift not in the same direction, but the counts get brighter from F070W to F200W and then fainter up to F444W?
Since we are looking at the rest frame, the reason has to be in the intrinsic luminosity of the galaxies at their relative wavelengths and that can be seen by the luminosity functions at $z=0$. 
We plot them in Fig.~\ref{fig:LFs_z0}. The luminosity functions at $z=0$ follow the same pattern as the counts, getting brighter from F070W to F200W and dimmer for the rest of the longer wavelength filters. This is something not related to the reference frame but to the intrinsic luminosity of galaxies. In fact, as it can be seen in Fig.~\ref{fig:SED} in the main text, the rest frame SED always peaks between filters F150W and F200W.
This intrinsic luminosity behaviour affects the location of the break in the observed counts at different wavelengths. 
We plot the apparent magnitude of the break as a function of the wavelength in Fig.~\ref{fig:break_IGL}.
Using the top panel of fig.~11 of \cite{windhorst23} we reported with a green line the apparent magnitude of the peak of the intergalactic light (IGL) at different wavelengths, showing that it follows the same trend. This suggests that both the location of the break in the number counts and the peak of the IGL are driven by the intrinsic luminosity of galaxies. However IGL is not related to the intensity of the break as the latter is the results of the change in reference frame, or in other words, to the change of the $k$-correction with redshift. \\
%is driven by the This indicates that the moving of the break of the power law in the galaxy counts is driven by the intrinsic change in the luminosity of galaxies at different wavelengths rather than a change in the reference frame.   

\begin{figure} 
\centering 
\begin{minipage}{0.4\textwidth} % Smaller image 
\includegraphics[width=\linewidth]{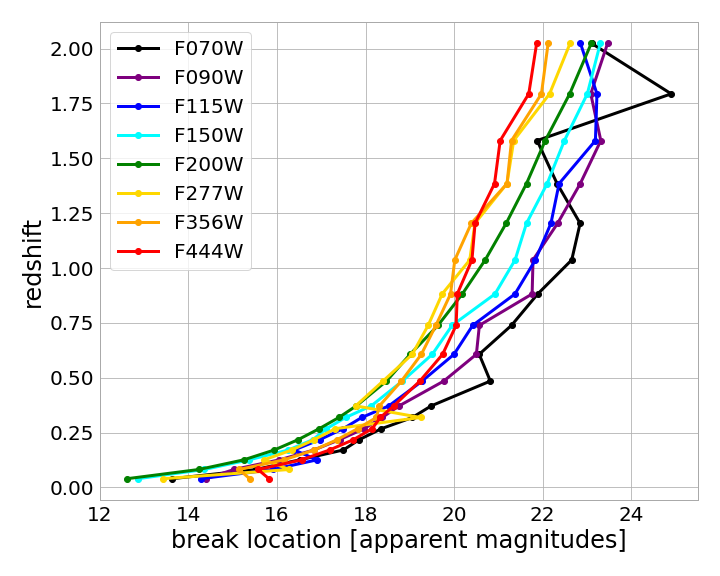} 
\end{minipage} 
\hspace{2mm} % Adjust spacing 
\begin{minipage}{0.56\textwidth} % Larger image 
\includegraphics[width=\linewidth]{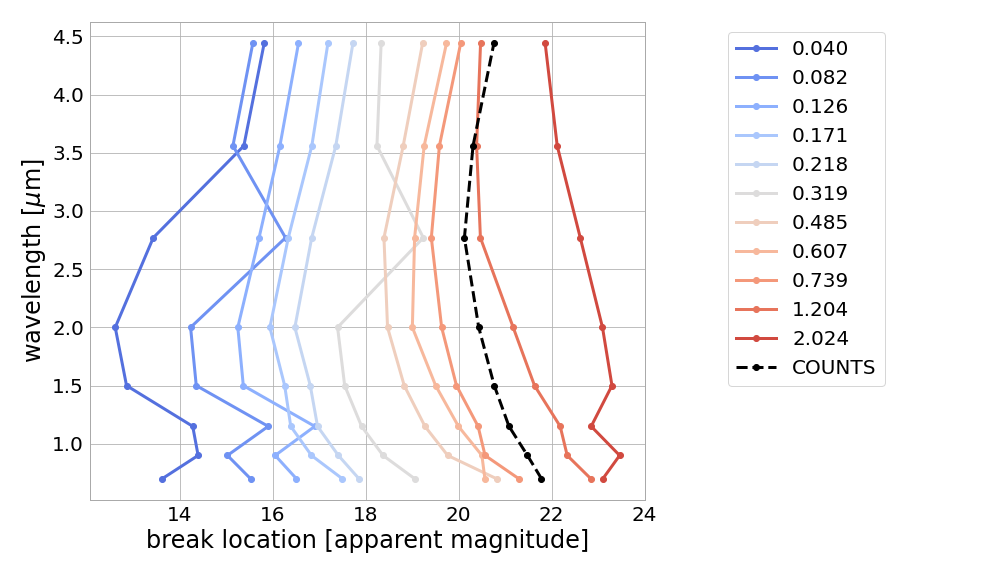} 
\end{minipage} 
\caption{Apparent magnitudes of the break in the power law for different luminosity functions as a function of redshift (left panel) and wavelength (right panel).} 
\label{fig:break_redshift}
\end{figure}

To fully exploit the power of the simulation, in Fig.~\ref{fig:break_redshift}, we compared the location of the break of luminosity functions at different redshift outputs with the location of the break in the galaxy number counts. 
This is helpful for identifying which redshift is dominating the apparent magnitude of the break in the number counts. 
In the left panel, we have the apparent magnitude of the break of the luminosity function as a function of redshift (colour-coded by the filter). 
From Fig.~\ref{fig:break_IGL}, we know that the break in the counts lies between $20-22$ apparent magnitudes. 
This is roughly covered by the luminosity functions between redshifts $0.75-1.00$. 
This could also be seen from the right panel of Fig.~\ref{fig:break_redshift} where we plot again the location of the break of the luminosity functions as a function of wavelength and colour-coded by the redshift. 
The location of the break of the number counts is drawn with a black-dashed line and it lies between the luminosity functions at $z=0.739$ and $z=1.204$, confirming that this is the redshift range dominating the number counts as outlined in the main text from the analysis of Figs.~\ref{fig:LF} and \ref{fig:redshift_filters}.

%\begin{figure}
%    \gridline{\fig{figs/fast5.png}{0.40\textwidth}{\hspace{-20mm}}
%                \fig{figs/fast2.png}{0.40\textwidth}{}}
%    \vspace{-10mm}
%    \gridline{\fig{figs/fast3.png}{0.40\textwidth}{\hspace{-20mm}}
%                \fig{figs/fast4.png}{0.40\textwidth}{}}
%    \caption{Comparison of the constituent luminosity functions for two representative filters, F444 (top row) and F090W (bottom row). The left column uses as $x$-axis the rest-frame absolute magnitudes while the right column is in the observer frame, by taking into account the effect of the k-correction.}
%    \label{fig:OBSREST_ABS}
%\end{figure}

%%%%%%%%%%%%%%%%%%%%%%%%%%%%%%%%%%%%
%%%%%%%%%%%%%%%%%%%%%%%%%%%%%%%%%%%%%%%%%%%%%%%%
%%%%%%%%%%%%%%%%%%%%%%%%%%%%%%%%%%%%%%%%%%%%%

\bibliography{manzoni}{}
\bibliographystyle{aasjournal}



\end{document}